\documentclass[12pt]{article}
\usepackage{amsmath,amssymb}
\usepackage{epsfig}
\usepackage{cite}

\def\m@th{\mathsurround=0pt }
\def\leftrightarrowfill{$\m@th \mathord\leftarrow \mkern-6mu
	\cleaders\hbox{$\mkern-2mu \mathord- \mkern-2mu$}\hfill
	\mkern-6mu \mathord\rightarrow$}

\def\overleftrightarrow#1{\vbox{\ialign{##\crcr
	\leftrightarrowfill\crcr\noalign{\kern-1pt\nointerlineskip}
	$\hfil\displaystyle{#1}\hfil$\crcr}}}

\newcommand{\tr}{{\rm Tr}}
\newcommand{\be}{\begin{equation}}
\newcommand{\ee}{\end{equation}}

\newcommand{\Tr}{\mathop{\rm Tr}}
\def\I{\rm 1\kern-.24em l}  

\def\shat{\ifmmode \hat{s}\else $\hat{s}$\fi}

\def\UV{UV}
\def\IR{IR}
\def\nn{\nonumber}

\newcommand{\newc}{\newcommand}

\newc{\gsim}{\lower.7ex\hbox{$\;\stackrel{\textstyle>}{\sim}\;$}}
\newc{\lsim}{\lower.7ex\hbox{$\;\stackrel{\textstyle<}{\sim}\;$}}
\newc{\ie}{{\it i.e.}}
\newc{\etal}{{\it et al.}}
\newc{\mev}{\hbox{\rm\,MeV}}
\newc{\gev}{\hbox{\rm\,GeV}}
\newc{\tev}{\hbox{\rm\,TeV}}
\newc{\xpb}{\hbox{\rm\, pb}}
\newc{\xfb}{\hbox{\rm\, fb}}

\newc{\G}{{\cal G}}
\newc{\h}{{\cal H}}
\newc{\D}{{\cal D}}
\newc{\E}{{\cal E}}
\newc{\x}{{\widehat x}}
\newc{\q}{{\widehat q}}

\newc{\mtop}{m_t}
\newc{\mbot}{m_b}
\newc{\mz}{M_Z}
\newc{\mw}{M_W}
\newc{\alphasmz}{\alpha_s(M_Z)}
\newc{\swsq}{\sin^2\theta_W}
\newc{\cwsq}{\cos^2\theta_W}
\newc{\tw}{\tan\theta_W}
\newc{\cw}{\cos\theta_W}
\newc{\sw}{\sin\theta_W}
\newc{\BR}{\hbox{\rm BR}}
\newc{\zbb}{Z\to b\bar}
\newc{\Gb}{\Gamma (Z\to b\bar b)}
\newc{\Gh}{\Gamma (Z\to \hbox{\rm hadrons})}
\newc{\sgn}{\mbox{sgn}}

\def\I{1\hspace{-4pt}1}
\def\ov{\overline}

\def\dd{\mathrm{d}}
\def\ii{\mathit{i}}

\newcounter{mysubequation}[equation]

\def\beq{\begin{equation}}
\def\eeq{\end{equation}}
\def\bea{\begin{eqnarray}}
\def\eea{\end{eqnarray}}
\def\slashchar#1{\setbox0=\hbox{$#1$}           
   \dimen0=\wd0                                 
   \setbox1=\hbox{/} \dimen1=\wd1               
   \ifdim\dimen0>\dimen1                        
      \rlap{\hbox to \dimen0{\hfil/\hfil}}      
      #1                                        
   \else                                        
      \rlap{\hbox to \dimen1{\hfil$#1$\hfil}}   
      /                                         
   \fi}                                         %
\catcode`@=11
\long\def\@caption#1[#2]#3{\par\addcontentsline{\csname
  ext@#1\endcsname}{#1}{\protect\numberline{\csname
  the#1\endcsname}{\ignorespaces #2}}\begingroup
    \small
    \@parboxrestore
    \@makecaption{\csname fnum@#1\endcsname}{\ignorespaces #3}\par
  \endgroup}
\catcode`@=12

\def\UV{\rm \textsc{uv}}
\def\IR{\rm \textsc{ir}}
\def\Le{{\bf L}}
\def\R{{\bf R}}
\def\l{{\bf l}}
\def\r{{\bf r}}
\def\A{{\bf A}}

\begin{document}

\baselineskip=18pt

\setcounter{footnote}{0}
\setcounter{figure}{0}
\setcounter{table}{0}

\begin{titlepage}
\begin{flushright}
\end{flushright}
\vspace{.3in}

\begin{center}
{\Large \bf
Massive Pions, Anomalies and Baryons \\[4pt]
in Holographic QCD
}
\\[6pt]
{\bf O.~Dom\`enech$^a$, G.~Panico$^{b}$ and A.~Wulzer$^{c}$}

\vspace{.5cm}

\centerline{$^{a}${\it Departament de F\'\i sica and IFAE, Universitat Aut\`onoma de Barcelona,
08193 Bellaterra, Barcelona}}
\centerline{$^{b}${\it Institute for Theoretical Physics, ETH Zurich,
8093 Zurich, Switzerland}}
\centerline{$^{c}${\it Institut de Th\'eorie des Ph\'enom\`enes Physiques, EPFL,
  CH--1015 Lausanne, Switzerland}}

\end{center}
\vspace{.8cm}

\begin{abstract}
\medskip
\noindent

We consider a holographic model of QCD, obtained by a very simple modification
of the original construction, which describes at the same time the pion mass,
the QCD anomalies and the baryons as topological solitons.
We study in detail its phenomenological implications in both the mesonic and baryonic
sectors and compare with the observations.

\end{abstract}

\bigskip
\bigskip

\end{titlepage}

\section{Introduction}

The virtues of 5d theories as phenomenological descriptions of the QCD hadrons in the
large-$N_c$ expansion have been widely discussed in the literature
\cite{Son:2003et,Erlich:2005qh,DaRold:2005zs,DaRold:2005vr,Hirn:2005nr,Pomarol:2007kr,Pomarol:2008aa,Panico:2008it} and can
be briefly summarized as follows. From the theoretical point of view, these models resemble
large-$N_c$ QCD in that they contain infinite towers of weakly interacting mesons while the baryons
are solitons, similar to the Skyrmions of the non-linear $\sigma$-model. Differently from ordinary
Skyrmions, however, the 5d soliton size remains finite in the weak coupling (large-$N_c$) limit
and is therefore parametrically larger than the length scale at which the 5d model enters
the strongly coupled regime. The 5d Skyrmions are genuinely macroscopic objects, whose physics
is perfectly within the reach of the effective theory as it must also be the case for the baryons
in the true theory of large-$N_c$ hadrons \cite{Witten:1979kh}.

For what concerns phenomenology, it is already quite remarkable that the holographic implementation
of the QCD chiral symmetry and of its breaking automatically leads to towers of  vector and scalar
mesons with the quantum numbers of the observed $\rho$, $\omega$, $a_1$, $f_1$, $a_0$, $f_0$, $\pi(1300)$
and $\eta(1295)$, but even more remarkable is that, when extrapolating the model to the physically
relevant case of $N_c=3$, an agreement  at less than $10\%$ with observations is found.
This success is non-trivial because the model contains an extremely limited number of free parameters
that can be adjusted to reproduce many observed meson couplings and masses, and is mainly due to a
set of phenomenologically successful ``sum rule'' relations among the predictions. The latter originate
from the 5d structure of the theory and are independent of many details of the model, such as
the background 5d metric, whose choice is therefore unessential for phenomenology as discussed
in \cite{Becciolini:2009fu}. The overall agreement is less good in the baryonic sector, but still
compatible with the expected $1/N_c\sim 30\%$ deviations. Furthermore, one recovers in a non-trivial way
several features of the large-$N_c$ QCD baryons such as the scaling of the currents form factors
with $N_c$, the fact that the isovector electric and magnetic radii diverge in the chiral limit
(as in QCD \cite{Beg:1973sc}) and other large-distance behaviors \cite{Cherman:2009gb}.
In holographic QCD, and in the present article, an important role is played by the Chern--Simons (CS)
term of the 5d action whose presence is required, with a fixed coefficient, by the need of reproducing
the QCD global anomalies. The CS is therefore the 5d analog of the gauged Wess--Zumino--Witten (WZW)
term of the standard 4d chiral theory and, like the WZW, it is responsible for the $\pi\rightarrow\gamma\gamma$
decay and for ``naive parity'' breaking in the Goldstone interactions \cite{Witten:1983tw}.
The CS actually contains the WZW, in the sense that it reduces to it in the low-energy
description of the Goldstones obtained from the 5d theory, but it also contains other interactions such
as the well-measured ``anomalous parity'' couplings $g_{\rho\pi\gamma}$, $g_{\omega\pi\gamma}$
and $g_{\omega\rho\pi}$. As a result of the 5d structure of the theory, these couplings are predicted
and are found to be in good agreement with the observations. In the case of exact chiral symmetry
(and only in that case, as we will show) the CS term is also crucial for the physics of the baryons
because it stabilizes the 5d Skyrmion solution and makes its size scale like a constant for large $N_c$.

Many of the above-mentioned results, and in particular the ones related with the CS term and with the baryons,
have however been only established in the minimal holographic QCD model with exact chiral symmetry;
the aim of this article is to generalize them to the more realistic case of explicit chiral breaking
(and therefore non-vanishing mass for the Goldstones) and to check if the previously outlined general picture survives.
In order to complete this program we first of all need a model with explicit breaking, but we cannot simply
employ the original one of \cite{Erlich:2005qh,DaRold:2005zs,DaRold:2005vr}. A very simple variation is needed
in order to incorporate the CS term and the 5d Skyrmions that were not considered in the original literature.
Like in \cite{Erlich:2005qh,DaRold:2005zs,DaRold:2005vr}, our model will be a 5d $U(2)_L\times U(2)_R$ gauge
theory with a 5d scalar field $\Phi$ in the bifundamental of the group, whose presence is necessary to parametrize
the breaking of the chiral symmetry due to the quark mass term.
The only difference with \cite{Erlich:2005qh,DaRold:2005zs,DaRold:2005vr}
is in the boundary conditions for the gauge fields at the IR brane, where we break the chiral group to vector
(as in \cite{Hirn:2005nr}) instead of preserving it and making the spontaneous chiral symmetry breaking arise
exclusively from the VEV of $\Phi$. 
\footnote{Actually, another difference
with \cite{Erlich:2005qh,DaRold:2005zs,DaRold:2005vr} 
is that the gauge group was taken there to be the chiral $SU(2)$ 
while our model is based on $U(2)$.}
This choice of boundary conditions would be a priori equivalent to the
original one from the model-building perspective, and costs no more parameters, but is necessary if willing
to incorporate the CS term. As we will discuss in more details in the following section (see also \cite{Panico:2007qd}),
the reason is that the CS gauge variation is an integral on the boundaries of the 5d space and therefore receives
contribution from both the UV and the IR brane. While the UV term is welcome because it reproduces
the QCD global anomalies, the IR one must cancel because the IR group is gauged and a non-vanishing variation
would have on the model the same effects of a gauge anomaly. The symmetry-breaking boundary conditions,
which are actually equivalent to gauging only the vector subgroup at the IR, enforce this cancellation.
On top of this, symmetry-preserving IR boundary conditions could make the 5d Skyrmion unstable or even
the classical 5d Skyrmion solution not to exist at all because the baryonic
charge \footnote{For the notation, see \cite{Pomarol:2007kr,Pomarol:2008aa,Panico:2008it} or the following Section.}
\bea
B&&\displaystyle{=\,\frac1{32\pi^2}\int d^3x\int^{z_{\IR}}_{z_{\UV}} dz\,
\epsilon_{\hat\mu\hat\nu\hat\rho\hat\sigma}\Tr\left[{{\bf L}^{\hat\mu\hat\nu}{\bf L}^{\hat\rho\hat\sigma}}
-{{\bf R}^{\hat\mu\hat\nu}{\bf R}^{\hat\rho\hat\sigma}}\right]\,}\nn\\
&&\displaystyle{\propto \,
\int_{r\rightarrow\infty}\hspace{-15pt}\left[\omega_3(l)-\omega_3(r)\right]\,+\,\int_{z=z_{\IR}}\hspace{-15pt}\left[\omega_3(l)-\omega_3(r)\right]\,
}\,,
\label{Bch}
\eea
would not be quantized and therefore not topologically conserved. It is indeed possible to change continuously
the IR contribution to $B$ (the last term in the equation above) without affecting the rest by a local variation
of the fields at the IR boundary. This is impossible with symmetry-breaking boundary conditions because the IR
contribution vanishes identically and the baryonic charge is quantized as shown in \cite{Pomarol:2007kr,Pomarol:2008aa,Panico:2008it}.

The paper is organized as follow. In Section~\ref{sec:MassivePions} we first of all describe the
model which is, to our knowledge, the first one containing at the same time the explicit chiral breaking,
the QCD anomalies ({\it{i.e.}} the CS term) and the 5d Skyrmion. We also discuss its phenomenological
implications for the physics of vector and scalar mesons, which because of the change in the boundary
conditions are a priori different from the ones derived in \cite{Erlich:2005qh,DaRold:2005zs,DaRold:2005vr}.
We actually find a quite similar phenomenology, the only remarkable difference being in the mass of
the $\pi(1300)$ pseudoscalar meson that was impossible to fit in the original model \cite{DaRold:2005vr}
while it is easily accommodated in our case. We obviously also take into account the processes mediated
by the CS term that were absent in the original model, and eventually perform a complete fit of
the meson observables which allows us to fix the parameters and to quantify the level of agreement with
observations. In Section~\ref{sec:Baryons} we study the physics of the baryons, generalizing the analysis
of \cite{Pomarol:2007kr,Pomarol:2008aa,Panico:2008it}
where the chiral case was considered. In that section we check that the existence and calculability
of the 5d Skyrmion is maintained in the non-chiral case, and see how much the predictions change after
the deformations required to incorporate the chiral breaking in the holographic model. The latter deformation
is not mild because the newly introduced 5d scalar does not decouple from the vector mesons, and therefore
from the 5d Skyrmion solution, even when the chiral symmetry is restored. We also compute the isovector
form factor radii which were divergent in the chiral case and become now finite, as expected in QCD,
due to the chiral breaking. Finally, in Section~\ref{sec:Conclusions}, we present our conclusions.

\section{Massive Pions and Scalar Resonances}\label{sec:MassivePions}

As a starting point for the description of the model, let us consider the QCD partition function $Z_{QCD}[l,r,M]$ in the presence of sources for the left and right global currents and for the quark mass operator. The latter reads
\be
\displaystyle{
Z_{QCD}\left[\l,\r,M\right]\,=\,\int {\mathcal{D}}\Psi \, {\rm{exp}}\left[iS_{QCD}[\Psi]\,+i \int d^4x\, {\rm{Tr}}\left(\l_\mu j_L^\mu+\r_\mu j_R^\mu-
 Ms - s^\dagger M^\dagger
\right)\right]\,,}
\label{def}
\ee
where $\Psi$ collectively indicates the QCD fundamental fields, $S_{QCD}$ is the massless QCD action, $\left(j^\mu_{L,R}\right)_{ij}=\overline{q}_{L,R}^j\gamma^\mu q_{L,R}^i$ are the currents of  the $U(2)_L\times U(2)_R$ chiral group and $s_{ij}=\overline{q}_L^j q_R^i$ is the quark bilinear. Due to anomalies, the partition function changes under local  chiral transformations as \footnote{The equation which follows is only valid in the large-$N_c$ limit in which the $U(1)_A$--$SU(N_c)^2$ anomaly can be neglected and the $\eta'$ only gets its mass from the explicit chiral symmetry breaking. Given that the expansion parameter of the 5d theory is interpreted as $1/N_c$, it is completely correct to stick to eq.~(\ref{an}) at the leading order, though it would be interesting to see what kind of next-to-leading order corrections the inclusion of the $U(1)_A$ anomaly leads to. The problem of incorporating the $\eta'$ mass in holographic QCD has already been addressed in \cite{Katz:2007tf}, though its connection with the anomalies and the $1/N_c$ expansion has not been outlined.\label{footnote anomaly}}
\be
\displaystyle{
{\displaystyle{Z_{QCD}\left[\l^{({\widehat{g}}_L)},\r^{({\widehat{g}}_R)},{\widehat{g}}_L M\, {\widehat{g}}^\dagger_R\right]\,=\,e^{i{\mathcal{A}}}Z_{QCD}\left[\l,\r,M\right]}}}\,,
\label{an}
\ee
where the vector sources transform as gauge fields ({\it{i.e.}}, $\l_\mu^{({\widehat{g}}_L)}={\widehat{g}}_L[\l_\mu+i\partial_\mu]{\widehat{g}}_L^\dagger$ and analogously for $\r_\mu$) and ${\mathcal{A}}$ stands for the QCD global anomaly which,
as discussed in appendix~\ref{app:Anomaly}, can be put in the form
\be
\mathcal{A} = \frac{N_c}{24 \pi^2} \int \left[\ov\omega_4^1({\boldsymbol \alpha}_L, {\bf l})
- \ov\omega_4^1({\boldsymbol \alpha}_R, {\bf r})\right]
\label{eq:Anomaly}
\ee
by the addition of suitable local counterterms.
In words, what eqs.~(\ref{def}) and (\ref{an}) mean is that 2-flavor (large-$N_c$) QCD is a theory endowed with an $U(2)_L\times U(2)_R$ anomalous global symmetry and with a non-dynamical spurion field $M$, associated with the quark mass operator, in the bifundamental of the group. This symmetry, with its explicit breaking parameter $M$, must obviously be present in any phenomenological description of hadrons such as the one we want to construct.

Following the holographic method, we incorporate the chiral symmetry and the spurion by introducing 5d fields with appropriate quantum numbers ($\Le_M$, $\R_M$ and $\Phi$) associated respectively to the sources $\l_\mu$, $\r_\mu$ and $M$, and identify the latter with the values of these fields at the UV-boundary. For the gauge fields the model is exactly as in \cite{Pomarol:2008aa}, and all what we have to add is the scalar $\Phi$ which, according to the holographic prescription, transforms as
\begin{equation}
{\displaystyle{\Phi \, \rightarrow \, \Phi^{(g_L,g_R)}\,\equiv\, g_L \Phi g_R^\dagger\,,}}
\end{equation}
under the local 5d $U(2)_L\times U(2)_R$ group and is subject to the following UV boundary conditions
\beq
\displaystyle{\Phi\left|_{z=z_{\UV}}\right.=\left(\frac{z_{\UV}}{z_{\IR}}\right)^{\Delta^-}M\, ,}
\label{uvboundary condition}
\eeq
with $\Delta^-$ defined in eq.~(\ref{delta definition}).
The need for the rescaling in $M$ is a technicality  associated with our choice of the AdS$_5$ geometry, and will be explained in the following. For simplicity, and since we are considering the two-flavor case in which the vector symmetry breaking is negligible, we take the spurion VEV to be of the form
$$
M\,=\, M_q\,\I\,,
$$
which respects $U(2)_V$.

The partition function of the 5d model, which we would like to identify with the QCD one in eq.~(\ref{def}), is defined as
\be
{\displaystyle{Z\left[\l,\r,M\right]\,\equiv\,\int \mathcal{D} L_{M}   \mathcal{D} R_M   \mathcal{D} \Phi \exp{\left\{i\,S_5[L,R,\Phi]\right\}
}
}}\,,
\label{pf}
\ee
where the dependence of the r.h.s. on the sources arises from the boundary conditions on the allowed field configurations. The gauge part of the 5d action is given by a kinetic
part
\begin{equation}
S_g=-\int d^4{x}\int^{z_{\IR}}_{z_{\UV}} dz\,  a(z)\, \frac{M_5}{2} \left\{
\Tr\left[{L_{MN}L^{MN}}\right]\,+\,\frac{1}2{\widehat{L}}_{MN}{\widehat{L}}^{MN}\,+\,\{L\,\leftrightarrow\,R\}\right\}\, ,
\label{Sg}
\end{equation}
and a Chern--Simons part
\begin{equation}
S_{CS}\,=\,\frac{N_c}{16\pi^2}\int d^5x\left\{
\frac14\epsilon^{MNOPQ}{\widehat{L}_M}\Tr\left[L_{NO}L_{PQ}\right]\,+\,
\frac1{24}\epsilon^{MNOPQ}{\widehat{L}_M}{\widehat{L}_{NO}}{\widehat{L}_{PQ}}\,-\,\{L\,\leftrightarrow\,R\}
\right\}\,,
\label{Scs}
\end{equation}
having parametrized the fields as ${\bf L}_\mu = L_\mu^a \sigma^a/2 + \widehat L_\mu \I/2$
and analogously for ${\bf R}_\mu$.
As discussed before, the UV boundary conditions for the gauge fields are
\begin{equation}
\Le_\mu\left|_{z=z_{\UV}}\right.=\,{\bf l}_\mu\ ,
\;\;\;\;\;\; \R_\mu\left|_{z=z_{\UV}}\right.=\,{\bf r}_\mu
\,.\label{UVgaugebc}
\end{equation}
The action for the scalar is
\begin{equation}
\displaystyle{
S_\Phi = M_5 \int d^4 x \int^{z_{\IR}}_{z_{\UV}} dz\, a^3(z)
\left\{
{\rm Tr}\left[\left(D_M \Phi\right)^\dagger D^M \Phi\right]
- a(z)^2 M^2_{Bulk} {\rm Tr}\left[\Phi^\dagger \Phi\right]\right\}\,,}
\label{sscalar}
\end{equation}
where we defined the covariant derivative
\begin{equation}
\displaystyle{D_M \Phi \equiv \partial_M \Phi - i\, \Le_M \Phi + i\, \Phi \R_M\,.}
\end{equation}
If a local 4d chiral transformation is performed on the sources, as in eq.~(\ref{an}), the UV boundary
conditions get modified but this change can be reabsorbed in a change of variable in the path integral
of the form of a local 5d gauge transformation $g_{L,R} = \exp(i\alpha_{L,R})$ which does not reduce to the identity at the UV brane.
The kinetic part of the gauge action $S_g$ and the scalar action $S_\Phi$ are invariant under this transformation,
while the CS part (as discussed in appendix~\ref{app:Anomaly}) gives the variation
\begin{equation}\label{eq:DeltaS}
\Delta_\alpha S_{CS} = \frac{N_c}{24 \pi^2} \left(\int_{z=z_{\UV}}\hspace{-15pt}\left[\ov\omega_4^1({\boldsymbol \alpha}_L, {\bf l})
- \ov\omega_4^1({\boldsymbol \alpha}_R, {\bf r})\right]
- \int_{z=z_{\IR}}\hspace{-15pt}\left[\ov\omega_4^1({\boldsymbol \alpha}_L, {\bf L})
- \ov\omega_4^1({\boldsymbol \alpha}_R, {\bf R})\right]\right)\,.
\end{equation}
The UV contribution in the above expression is welcome, because it coincides with the
anomalous variation in eq.~(\ref{an}). On the other hand, the IR terms in $\Delta_\alpha S_{CS}$ have no
counterpart in 4d QCD and must be cancelled. The chiral-breaking conditions
\begin{equation}
\left(\Le_\mu-\R_\mu\right)\left|_{z=z_{\IR}}\right.=0\ ,\;\;\;\;\;\;\left(\Le_{\mu 5}+\R_{\mu 5}\right)\left|_{z=z_{\IR}}\right.=0
\label{IRgaugebc}
\end{equation}
enforce this cancellation, which, on the contrary, would not occur in the case of the symmetry-preserving
Neumann conditions considered in~\cite{Erlich:2005qh,DaRold:2005vr,DaRold:2005zs}.
The choice (\ref{IRgaugebc}) of boundary conditions
is motivated not only by the need of correctly reproducing the QCD anomaly,
as the previous discussion shows, but also by the consistency of the theory
in the presence of the CS term. We can introduce a term in the action only if
it respects all the symmetries which are gauged.
As can be seen from es.~(\ref{eq:DeltaS}), the CS term in eq.~(\ref{Scs}) is not invariant under local
axial transformations which do not vanish at the IR boundary, thus it can be consistently
introduced only if the local axial symmetry at the IR boundary is not gauged, as 
implied 
by the conditions in eq.~(\ref{IRgaugebc}).

For the scalar we impose the IR boundary condition
\beq
\displaystyle{\Phi\left|_{z=z_{\IR}}\right.\,=\,\pm\,\xi\,\I\, ,}
\label{irboundary condition}
\eeq
where $\xi$ is a positive parameter and fields satisfying eq.~(\ref{irboundary condition}) with both signs,
associated to two disconnected sectors, coexist in the theory.
These boundary conditions (with their sign ambiguity) can be thought to originate
(as in \cite{Erlich:2005qh,DaRold:2005zs,DaRold:2005vr}) from an IR localized potential of the form
$V=\lambda \left(({\rm Re\,Tr[\Phi]})^2 - 4 \xi^2\right)^2 + \eta \left(2 {\rm Tr}[\Phi^\dagger \Phi]
- ({\rm Re\,Tr[\Phi]})^2\right)$. The latter would enforce eq.~(\ref{irboundary condition})
in the $\lambda, \eta \rightarrow \infty$ limit.
When dealing with small field fluctuations around the vacuum, the subtle sign ambiguity is irrelevant, because only one of the two sectors contains a stable stationary point and the dynamics entirely takes place around it. This is because of the $\Phi\rightarrow-\Phi$ symmetry of the lagrangian, which implies
that a simultaneous sign change of $M_q$ and of the boundary condition (\ref{irboundary condition}) is
unobservable. The pion mass squared is therefore $m_\pi^2\sim \pm\,M_q
\xi\,f(z_{IR}\xi)+{\mathcal{O}}(M_q^2)$ which, depending on the sign of $M_q$, can be positive or
negative implying that only one of the two sectors can be tachyon-free. For $M_q>0$ the vacuum is in the
``plus-sign'' sector and the physics of mesons, described by small fluctuations around the vacuum, is
totally insensitive to the presence of the other (minus-sign) sector. We will show in the following
section that the baryons live, on the contrary, in the minus-sign sectors so that the sign ambiguity
in eq.~(\ref{irboundary condition}) has to be maintained.

Having described the model, we can now study its phenomenological implications for the physics of mesons, with the aim of comparing it with the observations and of fixing its parameters. With respect to the massless case of \cite{Pomarol:2008aa}, three new parameters ($M_q$, $\xi$ and $M_{Bulk}$) have been introduced, but also new predictions can be extracted from the model. The scalar $\Phi$ describes scalars and pseudo-scalars with the isospin quantum numbers of, respectively, the $a_0(980)$, the $f_0(980)$, the $\pi(1300)$ and the $\eta(1295)$;
we will use their masses, plus of course the pion mass, to fit the new
parameters. \footnote{In our model the axial field $A_M$ gives rise to towers of resonances
with the quantum numbers of the $f_1$ and of the $\eta$ mesons.
These states are predicted to be degenerate in mass with the
corresponding mesons of the $a_1$ and the $\pi$ tower. Although the masses
predicted for the $f_1(1285)$ and for the $\eta(1295)$
are close to the experimental ones, we decided not to include these
predictions in the fit because, in principle, they
could receive sizable corrections when a proper treatment of the
$U(1)_A$ anomaly is included (see footnote~\ref{footnote anomaly}).}

To study the meson spectrum, we have first of all to find the vacuum configuration, which is given by the solution of the bulk equation of motion for $\Phi$
\begin{equation}
D_M \left(a^3(z) D^M \Phi\right) = - a^5(z) M_{Bulk}^2 \Phi\,,
\end{equation}
for vanishing 4d momenta and with the boundary conditions in eqs.~(\ref{uvboundary condition}) and (\ref{irboundary condition}). The result is
\begin{equation}\label{PhiVEV}
\langle \Phi \rangle = c_+ \left(\frac{z}{z_{\IR}}\right)^{\Delta^+}
+ c_- \left(\frac{z}{z_{\IR}}\right)^{\Delta^-}\,\equiv\,v(z)\I,
\end{equation}
with
\begin{equation}\label{Phicoeff}
\left\{
\begin{array}{l}
c_+ = \displaystyle\frac{z_{\IR}^{2 \alpha}}{z_{\IR}^{2 \alpha}-z_{\UV}^{2 \alpha}}\left(\xi - M_q\right) \I\,,\\
\rule{0pt}{2.25em}c_- = \displaystyle\frac{1}{z_{\IR}^{2 \alpha}-z_{\UV}^{2 \alpha}}\left(z_{\IR}^{2 \alpha} M_q - z_{\UV}^{2 \alpha} \xi\right) \I\,,
\end{array}
\right.
\end{equation}
where we normalized the warp factor as $a(z) = z_{\IR}/z$ and we defined
\begin{equation}\label{delta definition}
\Delta^{\pm} = 2 \pm \sqrt{4 + M_\phi^2} = 2 \pm \alpha\,,
\end{equation}
with $M_\phi^2 \equiv z_{\IR}^2 M_{Bulk}^2$.
To obtain a real value for $\Delta^{\pm}$ and to avoid a singular behavior of the
$z^{\Delta^-}$ piece in eq.~(\ref{PhiVEV}) at the UV boundary in the $z_{\UV} \rightarrow 0$ limit,
we must impose the constraint $-4 \leq M_\phi^2 \leq 0$.
Notice that it is only because of the rescaling in eq.~(\ref{uvboundary condition}) that
we obtain a finite VEV in the $z_{\UV} \rightarrow 0$ limit while keeping all the
other parameters finite. The rescaling has allowed us to define a finite quantity,
$M_q$, which controls the departure from the chiral limit.

\subsection{The Mesons Wavefunctions}

In order to study the properties of the mesonic sector, it is useful to rewrite the scalar
field as
\begin{equation}
\Phi = \left(v\, \I + S\right) e^{i P/v}\,,
\end{equation}
where $S$ and $P$ are Hermitian matrices of respectively scalar and pseudoscalar fields.
Under the unbroken $U(2)_V$ symmetry $S$ and $P$ can both be decomposed as ${\bf 1} + {\bf 3}$.
Form the boundary conditions on the $\Phi$ field in eqs.~(\ref{uvboundary condition}) and
(\ref{irboundary condition}) we immediately find the boundary conditions
for the $S$ and the $P$ fields
\begin{equation}\label{SP boundary conditions}
\left. S \right|_{z_{\UV}} = \left. S \right|_{z_{\IR}} =
\left. P \right|_{z_{\UV}} = \left. P \right|_{z_{\IR}} = 0\,.
\end{equation}

A convenient gauge fixing choice for studying the properties of the mesons is the
$R_\xi$ gauge which eliminates the mixing between the gauge bosons
$A_\mu$, $V_\mu$ and the scalars $A_5$, $P$.
This gauge is obtained by introducing in the Lagrangian the terms
\begin{eqnarray}
{\cal L}^V_{gf} &=& - \frac{2 M_5 a(z)}{\xi_V} {\rm Tr}\left[
\partial_\mu V^\mu - \frac{\xi_V}{a(z)} \partial_z(a(z) V_5)\right]^2\,,\\
{\cal L}^A_{gf} &=& - \frac{2 M_5 a(z)}{\xi_A} {\rm Tr}\left[
\partial_\mu A^\mu - \frac{\xi_A}{a(z)} \partial_z(a(z) A_5)
- \xi_A a^2(z) v(z) P\right]^2\,,
\end{eqnarray}
where $V_M = ({\bf L}_M + {\bf R}_M)/2$ and $A_M = ({\bf L}_M - {\bf R}_M)/2$.
The wavefunctions and the masses of the mesons can be determined by solving the
quadratic equations of motion for the 5d fields and imposing the appropriate
boundary conditions. The computation is analogous to the one described
in \cite{DaRold:2005zs,DaRold:2005vr}, so we will skip here most of the details.
The physical degrees of freedom are easily identified using the unitary gauge, which corresponds to the limit
$\xi_{A,V} \rightarrow \infty$. In this gauge the fields $A_5$, $V_5$ and $P$ satisfy
the constraints
\begin{equation}\label{unitary gauge}
\partial_z (a(z) V_5) = 0\,, \qquad\quad
P = -\frac{1}{a^3(z) v} \partial_z(a(z) A_5)\,,
\end{equation}
and the pion field can be identified with the lightest KK mode of $A_5$.

The vector mesons, namely the $\rho$ and the $\omega$ mesons and their resonances,
are described by the KK modes of the vector gauge field $V_\mu$.
In the unitary gauge the $V_5$ component of the vector gauge field is forced to vanish
due to the boundary condition $\left. V_5\right|_{z_{\IR},z_{\UV}} = 0$, while
the 4d components $V_\mu$ satisfy the bulk equation of motion
\begin{equation}
\partial_\nu \partial^\nu V_\mu - \partial_\mu \partial^\nu V_\nu
- \frac{1}{a(z)} \partial_z \left(a(z) \partial_z V_\mu\right)
 = 0\,,
\end{equation}
and the boundary conditions
\begin{equation}
\left. V_\mu \right|_{z_{\UV}} = 0\,, \qquad\quad
\left. \partial_z V_\mu \right|_{z_{\IR}} = 0\,.
\end{equation}
As expected, the equation of motion for the vector mesons is independent of the VEV of the
5d scalar $\Phi$, and the masses and the wavefunctions for these states are exactly the same
as in \cite{Pomarol:2008aa}.

The equation of motion for the scalar field $S$ is independent of $v$ as well
\begin{equation}\label{equation S}
\partial_\mu \partial^\mu S - \frac{1}{a^3(z)} \partial_z \left(a^3(z)
\partial_z S\right) + a^2(z) M_{Bulk}^2 S = 0\,,
\end{equation}
and the corresponding boundary conditions are given in eq.~(\ref{SP boundary conditions}).
The KK modes of the $S$ field are interpreted as the parity-even scalar mesons,
whose lightest states are the $a_0(980)$ and the $f_0(980)$.

On the other hand the axial fields, which are interpreted as the
$a_1(1260)$ and $f_1(1285)$ mesons and their heavier resonances,
are sensitive to the symmetry breaking induced by the
VEV of $\Phi$, as can be seen from the equation of motion
\begin{equation}
\partial_\nu \partial^\nu A_\mu - \partial_\mu \partial^\nu A_\nu
- \frac{1}{a(z)} \partial_z \left(a(z) \partial_z A_\mu\right)
+2 a^2(z) v^2(z) A_\mu  = 0\,.
\end{equation}
The boundary conditions are
\begin{equation}
\left. A_\mu \right|_{z_{\UV}} = \left. A_\mu \right|_{z_{\IR}} = 0\,.
\end{equation}

Finally, the pion and the other pseudoscalar mesons are described, for finite $\xi_A$, by the fields $A_5$ and $P$.
These fields mix in the quadratic Lagrangian and it is not totally straightforward to obtain their equations of
motion in the unitary gauge limit $\xi_A\rightarrow\infty$. If one simply removes the $P$ field from
the Lagrangian through the constraint in eq.~(\ref{unitary gauge}) and then varies the action
with respect to $A_5$, one obtains indeed the bulk equation of motion
\begin{equation}\label{bulk EOM A5}
{\cal D} \left[\partial_\mu \partial^\mu A_5 + 2 a^2(z) v^2(z) {\cal D} A_5\right]  = 0\,,
\end{equation}
which is of fourth order in the $z$ derivatives, having defined the differential operator
\begin{equation}
{\cal D} \equiv 1 - \partial_z \left(\frac{1}{2 a^3(z) v^2(z)}\partial_z a(z)\right)\,.
\end{equation}
From the boundary conditions on $P$ and from the gauge fixing constraint one also gets the boundary conditions
\begin{equation}\label{boundary conditions A5}
\left. \partial_z (a(z) A_5) \right|_{z_{\UV}} = \left. \partial_z (a(z) A_5) \right|_{z_{\IR}} = 0\,,
\end{equation}
which however are only two while four conditions would be needed for the fourth order bulk equation (\ref{bulk EOM A5})
to have a unique solution. The two remaining boundary conditions cannot be obtained
directly in the unitary gauge, they arise from the equations of motion at finite $\xi_A$
by carefully taking the $\xi_A\rightarrow \infty$ limit. From this procedure we recover the bulk
equation (\ref{bulk EOM A5}) and the boundary conditions in (\ref{boundary conditions A5}), plus the additional constraints
\begin{equation}\label{extra conditions A5}
\left.\partial_z \left\{a(z) \left[
\partial_\mu \partial^\mu + 2 a^2(z) v^2(z)
{\cal D}\right]A_5\right\}\right|_{z_{\UV},z_{\IR}} = 0\,.
\end{equation}
By the above equation one can prove that the bulk equation (\ref{bulk EOM A5}) can be replaced by the reduced second-order one
\begin{equation}
\partial_\mu \partial^\mu A_5 + 2 a^2(z) v^2(z) {\cal D} A_5 = 0\,,
\end{equation}
with the boundary conditions given in eq.~(\ref{boundary conditions A5}). The pseudoscalar spectrum, and in particular the pion, its lighter state, obviously depends on $v(z)$ and therefore on $M_q$ and $\xi$. We have checked that for positive $M_q$ (and $\xi$) all the squared masses are strictly positive, while for $M_q<0$ the lightest state is tachyonic as anticipated in the discussion below eq.~(\ref{irboundary condition}).

\subsection{Fit on the Meson Observables}\label{sec:fit}

Before performing the complete fit, an estimate of the five parameters of our model ($L \equiv z_{\IR} - z_{\UV}$,
$M_5$, $M_{Bulk}$, $\xi$ and $M_q$) can be obtained as follows.
The length of the compact dimension is fixed by the $\rho$ meson mass through the
relation $m_\rho \simeq 2.4/L$, from which we derive $L^{-1} \simeq 320\ {\rm MeV}$ \cite{DaRold:2005zs}.
The bulk mass parameter is determined by the mass of the $a_0(980)$ meson. The
relation between these two quantities can be derived by solving analytically
the equation of motion for $S$ (eq.~(\ref{equation S})) and then imposing the
corresponding boundary conditions. This leads to an approximate relation
\begin{equation}
m_{a_0(980)} \simeq 1.3 \frac{\Delta^+}{L}\,,
\end{equation}
which gives $\Delta^+ \simeq 2.4$ or, equivalently, $L^2 M_{Bulk}^2 \simeq -3.8$.
An estimate of the value of the $\xi$ parameter can be obtained from the values of the
masses of the $\pi(1300)$ and of the $a_1(1260)$ mesons, which we determine numerically
using the previously determined values for $L$ and $M_{Bulk}$.
\begin{figure}
\centering
\includegraphics[width=.45\textwidth]{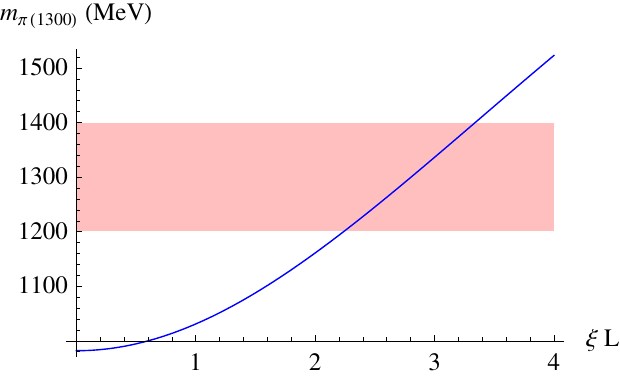}
\hspace{1.5em}
\includegraphics[width=.45\textwidth]{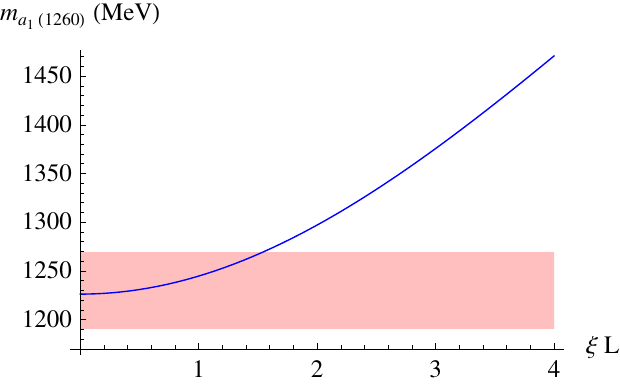}
\caption{Mass of the $\pi(1300)$ meson (left) and $a_1(1260)$ meson (right)
as a function of $\xi L$ for $L^{-1} = 320\ {\rm MeV}$ and $L^2 M_{Bulk}^2 \simeq -3.8$.
The shaded band shows the experimental values
 $m_{\pi(1300)} = (1300 \pm 100)\ {\rm MeV}$
and $m_{a_1(1260)} = (1230 \pm 40)\ {\rm MeV}$.}\label{xi fit}
\end{figure}
As can be seen from the plots in fig.~\ref{xi fit}, fitting the value of the $a_1(1260)$ mass
points towards the region of small $\xi$, $\xi L \lesssim 2$, while
reconstructing the $\pi(1300)$ mass favours larger value of the parameter
$2 \lesssim \xi L \lesssim 3$. Combining the two regions we get as a rough estimate
$\xi L \simeq 2$.
The values of $M_5$ and $M_q$ can be extracted respectively from the pion decay
constant $f_\pi$ and the pion mass $m_\pi$. These two quantities can be determined by
computing the holographic Lagrangian for the pion field as explained in appendix~\ref{app:HoloAction}.
We find that the series expansion for small $\xi L$ is given by
\begin{equation}\label{eq:fpi_mpi}
f_\pi^2 \simeq 4 \frac{M_5}{L} \left(1 + \frac{(\xi L)^2}{\Delta^+ ({\Delta^+}^2-1)}\right)\,,
\qquad \quad
m_\pi^2 \simeq 2 (\Delta^+ -2) \xi M_q\,.
\end{equation}
Using the experimental values \footnote{We did not include the
electroweak correction in the determination of the observables. For this reason we compare the
predictions of our model with the experimental value for pion decay constant with subtracted
electroweak contributions \cite{Holstein:1990ua} and with the mass of the $\pi^0$,
whose electroweak corrections are negligible \cite{Bijnens:1996kk}.}
$f_\pi^2 = 92\ {\rm MeV}$ and $m_{\pi} = 135\ {\rm MeV}$,
we get $M_5 L \simeq 0.015$ and $M_q \simeq 35\ {\rm MeV}$.

A more precise determination of the microscopic parameters can be obtained by performing
a fit on a larger set of well measured mesonic observables. This procedure provides also
a way to estimate the level of agreement of the model with the experimental results.
As a simple fitting procedure, we chose to minimize the root mean square error (RMSE)
of our predictions with respect to the experimental data (see \cite{Becciolini:2009fu}
for details on the fitting procedure).
We remark that for our analysis we take into account only the deviation of the theoretical
predictions of our model from the central value of the experimental results.
A more refined procedure, which however would be beyond the scope of our work,
should also take into account the experimental error with which the various
observables have been measured.

\begin{table}
\centering
\begin{tabular}{l@{\hspace{.275em}}l@{\hspace{.275em}}l@{\hspace{.275em}}c}
\hline
& Experiment & AdS$_5$ & Deviat.\\
\hline
$m_\pi$ & $135\ {\rm MeV}$ & $134\ {\rm MeV}$ & $0.6\%$\\
$m_{\pi(1300)}$ & $1300\ {\rm MeV}$ & $1230\ {\rm MeV}$ & $5.6\%$\\
$m_\rho$ & $775\ {\rm MeV}$ & $783\ {\rm MeV}$ & $1.0\%$\\
$m_\omega$ & $782\ {\rm MeV}$ & $783\ {\rm MeV}$ & $0.1\%$\\
$m_{a_1(1260)}$ & $1230\ {\rm MeV}$ & $1320\ {\rm MeV}$ & $7.6\%$\\
$m_{a_0(980)}$ & $980\ {\rm MeV}$ & $1040\ {\rm MeV}$ & $6.5\%$\\
$m_{f_0(980)}$ & $980\ {\rm MeV}$ & $1040\ {\rm MeV}$ & $6.5\%$\\
$f_\pi$ & $92\ {\rm MeV}$ & $89\ {\rm MeV}$ & $3.6\%$\\
$f_\rho$ & $153\ {\rm MeV}$ & $149\ {\rm MeV}$ & $2.7\%$\\
$f_\omega$ & $140\ {\rm MeV}$ & $149\ {\rm MeV}$ & $6.4\%$\\
$g_{\rho\pi\pi}$ & $6.0$ & $4.89$ & $22.7\%$\\
$g_{\omega\pi\gamma}$ & $0.72$ & $0.71$ & $1.1\%$\\
$g_{\rho\pi\gamma}$ & $0.22$ & $0.24$ & $7.9\%$\\
$g_{\omega\rho\pi}$ & $15.0$ & $15.6$ & $3.7\%$\\
\hline
\hline
\rule{0pt}{1.em}RMSE & & & $7.7\%$\\
\hline
\end{tabular}
\hspace{1.5em}
\begin{tabular}{l@{\hspace{.275em}}l@{\hspace{.275em}}l@{\hspace{.275em}}c}
\hline
& Experiment & AdS$_5$ & Deviat.\\
\hline
$m_\pi$ & $135\ {\rm MeV}$ & $133\ {\rm MeV}$ & $1.2\%$\\
$m_{\pi(1300)}$ & $1300\ {\rm MeV}$ & $1207\ {\rm MeV}$ & $7.7\%$\\
$m_\rho$ & $775\ {\rm MeV}$ & $842\ {\rm MeV}$ & $8.6\%$\\
$m_\omega$ & $782\ {\rm MeV}$ & $842\ {\rm MeV}$ & $7.7\%$\\
$m_{a_1(1260)}$ & $1230\ {\rm MeV}$ & $1387\ {\rm MeV}$ & $12.7\%$\\
$m_{a_0(980)}$ & $980\ {\rm MeV}$ & $1099\ {\rm MeV}$ & $12.1\%$\\
$m_{f_0(980)}$ & $980\ {\rm MeV}$ & $1099\ {\rm MeV}$ & $12.1\%$\\
$f_\pi$ & $92\ {\rm MeV}$ & $89\ {\rm MeV}$ & $3.1\%$\\
$f_\rho$ & $153\ {\rm MeV}$ & $158\ {\rm MeV}$ & $3.1\%$\\
$f_\omega$ & $140\ {\rm MeV}$ & $158\ {\rm MeV}$ & $12.7\%$\\
$g_{\rho\pi\pi}$ & $6.0$ & $5.33$ & $12.5\%$\\
$g_{\omega\pi\gamma}$ & $0.72$ & $0.74$ & $3.2\%$\\
$g_{\rho\pi\gamma}$ & $0.22$ & $0.25$ & $12.6\%$\\
$g_{\omega\rho\pi}$ & $15.0$ & $16.8$ & $11.8\%$\\
\hline
\hline
\rule{0pt}{1.em}RMSE & & & $9.6\%$\\
\hline
\end{tabular}
\caption{Meson observables used for the fit of the microscopic parameters.
The table on the left shows the results of a fit which minimizes the
overall root mean square error (shown in the last line). The values of the parameters obtained
in this way are $L^{-1} = 325\ {\rm MeV}$, $M_5 L = 0.014$, $L^2 M^2_{Bulk} = -3.7$,
$\xi L = 2.1$ and $M_q = 31\ {\rm MeV}$.
The table on the right gives the list of results obtained by minimizing the largest
deviation from the experiments. With this procedure we get
$L^{-1} = 350\ {\rm MeV}$, $M_5 L = 0.014$, $L^2 M^2_{Bulk} = -3.8$, $\xi L = 1.5$ and
$M_q = 40\ {\rm MeV}$.
}\label{fit results}
\end{table}
The list of observables used in the fit includes the masses of the lightest mesonic resonances
as well as some of their decay constants and couplings. The predictions of the model
and the experimental values are shown in the list on the left of table~\ref{fit results}.
We found that the best agreement with the data is obtained for the following
values of the parameters $L^{-1} = 325\ {\rm MeV}$, $M_5 L = 0.014$,
$L^2 M^2_{Bulk} = -3.7$, $\xi L = 2.1$ and $M_q = 31\ {\rm MeV}$,
which are close to the previously estimated ones.
The overall agreement with the experimental data is quite good
and almost all the observables show a deviation from the experimental values
smaller than $8\%$, resulting in a RMSE of $7.7\%$. A somewhat surprising
result of the fit is the fact that only one observable, namely the $g_{\rho\pi\pi}$
coupling seems to have a large deviation from the experiments. This deviation
is however not completely unexpected. In the class of models we are considering
one gets an approximate tree-level relation \cite{DaRold:2005zs}
\begin{equation}\label{modifiedKSRF}
m_\rho^2 \simeq 3 f_\pi^2 g_{\rho\pi\pi}^2\,,
\end{equation}
which differ by a factor $2/3$ form the experimentally well satisfied KSRF relation
$m_\rho^2 \simeq 2 f_\pi^2 g_{\rho\pi\pi}^2$. In our fit the predictions for the
pion decay constant and for the $\rho$ meson mass are in excellent agreement with the
data, thus the $g_{\rho\pi\pi}$ coupling must necessarily deviate from
the experiments in order for the relation (\ref{modifiedKSRF}) to be valid.

It is interesting to notice that the change in the IR boundary conditions for the gauge fields
with respect to the original model of~\cite{DaRold:2005vr} has some relevant consequences
on the predictions of the theory. In the original set-up the mass of the $\pi(1300)$ resonance
could not be reproduced and the first resonance of the axial gauge field was identified
with the $\pi(1800)$ state. In the present model, on the contrary, the $\pi(1300)$ meson
can be naturally accomodated.

To check the stability of our predictions, we can compare the previous results
with the ones obtained by using an alternative fitting procedure.
For this purpose, we chose to minimize the largest deviation of our predictions from the experiments.
In this way we obtained the list of result shown in the right panel of table~\ref{fit results}.
The deviations from the experiments are now more uniformly spread among the
various observables, with a maximal deviation of $\sim 13\%$. The overall RMSE is
$9.6\%$, which is still reasonable and only slightly higher than the one found in the
previous fit. The corresponding values of the microscopic parameters are
$L^{-1} = 350\ {\rm MeV}$, $M_5 L = 0.014$, $L^2 M^2_{Bulk} = -3.8$, $\xi L = 1.5$ and
$M_q = 40\ {\rm MeV}$, which are in good agreement with the previous determination.

The heavier resonances, which have not been included in the fits, show larger deviations
from the experimental values. For example the predicted mass for the $\pi(1800)$
is $2140\ {\rm MeV}$ with an $18\%$ deviation and for the $a_0(1450)$ it is $2070\ {\rm MeV}$
with a $40\%$ deviation. We remark, however, that the heavy resonances, being close
to the cut-off of the theory, are expected to have larger theoretical uncertainties.

\begin{table}
\centering
\begin{tabular}{l@{\hspace{2.75em}}r@{\hspace{2.75em}}r}
\hline
& Experiment & AdS$_5$\\
\hline
$L_1$ & $0.4 \pm 0.3$ & $0.47$\\
$L_2$ & $1.4 \pm 0.3$ & $0.95$\\
$L_3$ & $-3.5 \pm 1.1$ & $-2.8$\\
$L_4$ & $-0.3 \pm 0.5$ & $0.0$\\
$L_5$ & $1.4 \pm 0.5$ & $0.72$\\
$L_6$ & $-0.2 \pm 0.3$ & $0.0$\\
$L_8$ & $0.9 \pm 0.3$ & $0.45$\\
$L_9$ & $6.9 \pm 0.7$ & $6.0$\\
$L_{10}$ & $-5.5 \pm 0.7$ & $-6.0$\\
\hline
\end{tabular}
\hspace{2em}
\begin{tabular}{l@{\hspace{2.75em}}r@{\hspace{2.75em}}r}
\hline
& Experiment & AdS$_5$\\
\hline
$L_1$ & $0.4 \pm 0.3$ & $0.43$\\
$L_2$ & $1.4 \pm 0.3$ & $0.87$\\
$L_3$ & $-3.5 \pm 1.1$ & $-2.4$\\
$L_4$ & $-0.3 \pm 0.5$ & $0.0$\\
$L_5$ & $1.4 \pm 0.5$ & $0.68$\\
$L_6$ & $-0.2 \pm 0.3$ & $0.0$\\
$L_8$ & $0.9 \pm 0.3$ & $0.39$\\
$L_9$ & $6.9 \pm 0.7$ & $5.6$\\
$L_{10}$ & $-5.5 \pm 0.7$ & $-5.6$\\
\hline
\end{tabular}
\caption{Predictions for the coefficients of the ${\cal O} (p^4)$
terms in the ${\bf \chi PT}$ Lagrangian compared with the
experimental values \cite{Pich:1998xt}. The values are given in units
of $10^{-3}$. The microscopic parameters are fixed by
the fits on the observables in table~\ref{fit results}
(RMSE fit for the left table and `maximal deviation' fit for the
right table).}\label{L coefficients}
\end{table}
Other quantities that we can extract from the model
are the coefficients of the ${\cal O}(p^4)$ terms
in the ${\bf \chi PT}$ Lagrangian, which describe the interactions of the pions
with the left and right sources and with the spurion field related to the
quark masses. The computation can be performed by following the holographic
procedure outlined in appendix~\ref{app:HoloAction}.
Due to the non negligible experimental uncertainty, we decided not to include these
observables in the fit for the microscopic parameters. We also excluded from
the computation the $L_7$ coefficient which arises from integrating out
the Goldstone boson singlet related to the $U(1)_A$ anomaly, whose
complete treatment is not included in the present model
(see footnote~\ref{footnote anomaly}).
In table~\ref{L coefficients} we listed the predictions of the model for two sets
of microscopic parameters found with the RMSE and the `maximal deviation' fit.
The numerical results in the two cases are quite similar and show a good
agreement with the experimental data. The reduced $\chi^2$ for the RMSE fit
is $1.0$, while for the `maximal deviation' fit it is $1.2$, and the
deviations from the experimental data are always below $\sim 1.5\, \sigma$.

In the chiral-symmetric models without a bulk scalar field some
phenomenologically successful relations were found among the
coefficients of the ${\bf \chi PT}$ Lagrangian \footnote{See for example~\cite{Hirn:2005nr}.}:
\begin{equation}\label{coeff_rel}
L_2 = 2 L_1\,,\qquad L_9 = - L_{10}\,,\qquad L_4 = L_6 = 0\,,\qquad L_3 = - 6 L_1\,.
\end{equation}
In our set-up all these relations remain valid, except for the last
one $L_3 = - 6 L_1$, which receives some corrections but is still
well satisfied (compare~\cite{DaRold:2005zs, DaRold:2005vr}).
Notice that the first and third relations in eq.~(\ref{coeff_rel}), which are implied
by the large $N_c$ limit of QCD \cite{Gasser:1984gg}, are not modified in our model.

\section{Baryons From 5d Skyrmions}\label{sec:Baryons}

\subsection{The Static Solution}\label{sec:staticsolution}

In the present model, baryons arise as 5d Skyrmions and studying their properties requires
a slight modification of the methods of \cite{Pomarol:2007kr, Pomarol:2008aa, Panico:2008it},
where the massless case has been considered.
As a first step we will consider the static soliton configuration.
A convenient and automatically consistent 2d ansatz is obtained, as in the massless case,
by imposing the solution to be invariant under a certain set of symmetry transformations.
These are cylindrical symmetry ({\it{i.e.}}, the simultaneous action of $3$-space and $SU(2)_V$ rotations),
3d parity and time-inversion, defined as a change of sign of all the temporal components combined
with ${\widehat L}\rightarrow - {\widehat L}$ and ${\widehat R}\rightarrow - {\widehat R}$.
This leads to the following ansatz for the gauge fields
\be
\left\{
\begin{array}{l}
\displaystyle
{\ov R}^a_j({\bf x},z) = \displaystyle A_1(r,z) \x_a \x_j + \frac1r \varepsilon_{ajk} \x_k
-\frac{\phi_{(x)}}r\varepsilon^{(x,y)}\Delta^{(y),aj} \,,\\
\displaystyle
{\ov R}^a_5({\bf x},z) = \displaystyle  A_2(r,z)\x^a \,,\\
\displaystyle
\widehat {\ov R}_0({\bf x},z) = \displaystyle \frac{s(r,z)}r \,,
\end{array}
\right.
\label{sts}
\ee
where $r^2=\sum_i x^i x^i$, \  $\x^i=x^i/r$, $\varepsilon^{(x,y)}$ is the antisymmetric tensor with
$\varepsilon^{(1,2)}=1$ and the ``doublet'' tensors $\Delta^{(1,2)}$ are
\be
\Delta^{(x),ab}\,=\, \left[
\begin{array}{l}
 \epsilon^{abc}\x^c\\
\x^a\x^b - \delta^{ab}
\end{array}
\right]\,.
\label{defde}
\ee
Because of parity, eq.~(\ref{sts}) also determines the ansatz for the $\Le$ fields which
are given by $L_i({\bf x},z)=-R_i(-{\bf x},z)$, $L_{5,0}({\bf  x},z)=R_{5,0}(-{\bf x},z)$
and analogously for ${\widehat L}$, ${\widehat  R}$.

The ansatz for $\Phi$ is obviously obtained by imposing the same symmetries: cylindrical symmetry implies
\begin{equation}
\overline{\Phi}({\bf x}, z) = \omega_0(r,z) \frac{\I}{2} + i \omega_a(r,z) \frac{\sigma^a}{2}\,,
\label{anphi}
\end{equation}
where
\begin{equation}
\left\{
\begin{array}{l}
\omega_0(r,z) = \lambda_2(r,z) \,,\\
\rule{0pt}{1.5em}\omega_a(r,z) = \lambda_1(r,z) \,  \widehat x_a\,.
\end{array}
\right.
\end{equation}
It is easy to check that parity acts on $\Phi$ as $\Phi({\bf x},z,t)\rightarrow\Phi^\dagger(-{\bf x},z,t)$,
while under time-inversion we have $\Phi({\bf x},z,t)\rightarrow\sigma_2\Phi^*({\bf x},z,-t)\sigma_2$.
Imposing $\Phi$ to be invariant under these transformations simply implies that $\lambda_{1,2}$ are real.
It is useful to note that our ansatz preserves, again as in the massless case, a local $U(1)$ subgroup of
the original 5d chiral group corresponding to gauge transformations of the form $g_R= g$ and $g_L= g^\dagger$ with
\be
g = \exp[i \alpha(r,z) x^a \sigma_a/(2r)]\,.
\label{eq:resU1SU2}
\ee
Under this residual $U(1)$ the fields $\phi=\phi_1+i\,\phi_2$, $s$ and $A_{\overline{\mu}}$ in eq.~(\ref{sts})
are respectively one charged and one neutral scalar and a gauge field. It is easy to check that the field
$\lambda=\lambda_1+i\,\lambda_2$ in eq.~(\ref{anphi}) also transforms as a charge-one scalar;
it will be convenient to define its 2d covariant derivative as
\begin{equation}
D_{\bar \mu} \lambda \equiv \partial_{\bar \mu} \lambda - i A_{\bar \mu} \lambda\,.
\end{equation}

It is straightforward to plug the ansatz in the 5d lagrangian and to obtain the energy of the Skyrmion.
From the gauge part in eq.~(\ref{Sg}) and (\ref{Scs})
we obtain, as in \cite{Pomarol:2008aa,Panico:2008it},
\be
\begin{array}{l}
\displaystyle
E_G=8\pi M_5\int_0^\infty dr\int^{z_{\IR}}_{z_{\UV}}dz\,\left\{a(z)\left[
|D_{\bar\mu}\phi|^2+\frac{1}{4}r^2 A^{2}_{\bar\mu\bar\nu}
+\frac{1}{2r^2}\left(1-|\phi|^2\right)^2-\frac12\left(\partial_{\bar\mu} s\right)^2\right]\right.\\
\displaystyle\left.
-\frac{\gamma L}2
\frac{s}{r}\epsilon^{\bar\mu\bar\nu}
\bigg[\partial_{\bar\mu}(-i \phi^*D_{\bar\nu}\phi+h.c.)
+A_{\bar\mu\bar\nu}\bigg]
\right\}
\,,
\end{array}
\label{eq:mass}
\ee
where
\begin{equation}
\gamma \equiv \frac{N_c}{16 \pi^2 M_5 L}\,,
\end{equation}
while the new contribution coming from the scalar part in eq.~(\ref{sscalar}) is
\begin{equation}
E_\Phi = 8 \pi M_5 \int dr \int dz \left\{
a^3(z)\left[\frac{r^2}{4}(D_{\bar\mu}\lambda)^* (D_{\bar\mu} \lambda)
-\frac{1}{8}(\phi \lambda^* - \lambda \phi^*)^2\right]
+ a^5(z) \frac{r^2}{4} M^2_{Bulk} \lambda^* \lambda\right\}\,.
\label{Ephi}
\end{equation}
Notice that the total energy $E=E_G+E_\Phi$ does not yet give the Skyrmion mass because
the infinite  energy of the vacuum needs to be subtracted in order to get an observable
quantity. This zero-point energy is obtained from eq.~(\ref{Ephi}) by plugging in the
vacuum field configuration which is given by
\be
\lambda\,=\,\lambda_{V}\,=\, 2\,i\,v(z)\,,\quad \phi\,=\,\phi_{V}\,=-\,i\,,
\label{vac}
\ee
all other fields vanishing.

The 2d EOM for the Skyrmion are easily derived, at this point, by varying the energy in eq.~(\ref{eq:mass},\ref{Ephi}), but in order for them to be solved suitable boundary conditions need to be specified at the four boundaries ($z=z_{\IR}$, $z=z_{\UV}$, $r=0$ and $r\rightarrow\infty$) of our 2d space. At $z=z_{\UV}$ and $z=z_{\IR}$ the boundary conditions are given, up to the sign ambiguity in eq.~(\ref{irboundary condition}) that we will now fix, by the ones discussed in the previous section. The presence of the $r=0$ boundary merely results from a choice of coordinates, the physical 5d space being completely regular at $r=0$. The boundary conditions will therefore result from just imposing regularity of the 5d fields, with no need for extra assumptions. At $r\rightarrow\infty$ we must require that the solution will have a finite mass, and this is ensured by imposing it to reduce, up to a symmetry transformation, to the vacuum configuration in eq.~(\ref{vac}). We also want a $B=1$ solution, where the baryon charge $B$ is defined in eq.~(\ref{Bch}) and is given by
\beq
B=\frac{1}{2\pi}\int_{0}^\infty dr\int^{z_{\IR}}_{z_{\UV}}dz\,\epsilon^{\bar\mu\bar\nu}
\bigg[\partial_{\bar\mu}(-i \phi^*D_{\bar\nu}\phi+h.c.)
+F_{\bar\mu\bar\nu}\bigg]\,,
\label{topcharge2d}
\eeq
in terms of the 2d fields. The above equation can be easily rewritten (as it must, being the topological charge) as a 1d integral on the boundaries of the 2d space, and in order to get $B=1$ from the  $r\rightarrow\infty$ boundary we must have, as in the massless case, the following boundary conditions
\be
\displaystyle{
r\rightarrow\infty\ :
\quad
\left\{
\displaystyle{
\begin{array}{l}
\displaystyle{
\phi = -i e^{i \pi (z-z_{\UV}) /(z_{\IR}-z_{\UV})}}\\
\displaystyle{
A_2 = \frac{\pi}{(z_{\IR}-z_{\UV})}}\\
\displaystyle{
s=0}
\end{array}}
\right.\,,}
\ee
which are obtained from the vacuum (\ref{vac}) by means of a residual $U(1)$ transformation in the form of eq.~(\ref{eq:resU1SU2}), with $\alpha=\pi (z-z_{\UV}) /(z_{\IR}-z_{\UV})$.

Consistently, the boundary conditions for $\lambda$ are obtained in the same way and read
\be
r\rightarrow\infty\ :
\quad
\lambda\, =\, i\, e^{i \pi (z-z_{\UV}) /(z_{\IR}-z_{\UV})}\,2\, v(z)\,.
\ee
At $z=z_{\IR}$, the above equation implies $\lambda(r=0,z=z_{\IR})= -2\,i\,\xi$, because the ``twist'' $e^{i \pi (z-z_{\UV})/(z_{\IR}-z_{\UV})}$ reduces to $-1$ at the IR while the vacuum respects the boundary condition (\ref{irboundary condition}) with the plus sign. This resolves the ambiguity and enforces the 5d Skyrmion to live in the minus-sign sector, with IR boundary conditions given by
\be
z=z_{\IR}\ :
\quad
\lambda\, =\, -2\,i\,\xi \,.
\ee
We stress, as mentioned in the discussion below eq.~(\ref{irboundary condition}), that the sign
ambiguity in the IR boundary conditions results from our choice of giving generalized Dirichlet
conditions on $\Phi$, instead of treating it as a Neumann field and making its boundary
conditions originate from a localized potential that would cost us more new parameters.
If we had made the other choice, we would have had no ambiguity, and consequently no separated
sectors in the field space. If studying this different setup in the limit of infinite strength
for the coupling in the localized potential we expect that, while the vacuum and the meson's
wave function will be found to fulfill eq.~(\ref{irboundary condition}) with the plus sign,
the other boundary condition will be enforced on the Skyrmion solution and the results of the
present paper will be recovered. Coming back to the boundary conditions, we must still specify
the ones at $r=0$ and at $z=z_{\UV}$. These are
\be
r\rightarrow0\ :
\quad
\left\{\begin{array}{l}
\displaystyle{\lambda_1\, =\, 0}\\
\displaystyle{\partial_r\lambda_2\, =\, 0}
\end{array}
\right.
\quad\qquad
z=z_{\UV}\ :
\quad
\displaystyle{
\lambda\, =\, 2\,i\,\left(\frac{z_{\UV}}{z_{\IR}}\right)^{\Delta^-}\, M_q}
\ee
where the ones for $r=0$ arise, respectively, from asking the 5d field $\overline{\Phi}$ and its 3-space derivative to be regular and single-valued. For all the other fields the boundary conditions are the same of the massless case and are reported in appendix~\ref{app:EOM}.

\subsection{Zero-Mode Fluctuations}

In order to describe the baryons we need to consider the time-dependent deformations of
the static soliton solution.
The analysis proceeds exactly as in the massless case, we will therefore skip most of
the details and refer the interested reader to ref.~\cite{Panico:2008it}.

The single-baryon states can be identified with the
zero-mode fluctuations, thus an analysis of the infinitesimal deformations will be
sufficient for our purposes. The relevant configurations
are the ones which describe a slowly-rotating solution, whose degrees of freedom
can be parametrized by three collective coordinates encoded in an $SU(2)$ matrix
$U(t)$.

To describe the solution we need to generalize the ansatze given in eqs.~(\ref{sts})
and (\ref{anphi}). The ansatz for the gauge fields
is analogous to the one for the massless pion case:
\be
\displaystyle
R_{\hat\mu}({\bf x},z;U)\,=\,U\,{\ov R}_{\hat\mu}({\bf x},z)\,U^\dagger\,,
\;\;\;\;\;\;\;\;\;
{\widehat{R}_{0}}({\bf x},z;U)\,=\,{{{\widehat{\ov{R}}}}}_{0}({\bf x},z)\,,
\label{ans0}
\ee
and
\be
\displaystyle
R_{0}({\bf x},z;U)\,=\,U\,{\ov R}_{0}({\bf x},z;K)\,U^\dagger\,+\,i\,U\partial_0U^\dagger \, ,
\;\;\;\;\;\;\;\;\;
{\widehat{R}_{\hat\mu}}({\bf x},z;U)\,=\,{{{\widehat{\ov{R}}}}}_{\hat\mu}({\bf x},z;K)\,,
\label{ansk}
\ee
where
\be
\left\{
\begin{array}{l}
{\ov R}_{0}^a({\bf x},z;K) =\displaystyle \chi_{(x)}(r,z)k_b\Delta^{(x),ab} + w(r,z) (k\cdot\x)\x^a\,,\\
\displaystyle
{{{\widehat{\ov{R}}}}}_{i}({\bf x},z;K) = \displaystyle \frac{\rho(r,z)}{r}\left(k^i - (k\cdot\x)\x^i\right) + B_1(r,z)(k\cdot\x)\x^i + Q(r,z)\epsilon^{ibc}k_b\x_c \,,\\
\displaystyle
\rule{0pt}{1.5em}{{{\widehat{\ov{R}}}}}_{5}({\bf x},z;K) = \displaystyle B_2(r,z)(k\cdot\x) \,.
\end{array}
\right.
\label{ansk1}
\ee
The ansatz for the scalar field $\Phi$ is given by
\be
\Phi({\bf{x}},\,z;\,U)\,=\, U{\overline{\Phi}}({\bf{x}},\,z)\,U^\dagger\,,
\label{ansphi}
\ee
where ${\overline{\Phi}}$ is as in eq.~(\ref{anphi}) with the new definitions
\begin{equation}
\left\{
\begin{array}{l}
\omega_0(r,z) = \exp\left[-i (k\cdot \widehat x) \rho(r,z)\right]
\left(\lambda_2(r,z) + i (k\cdot \widehat x) \eta_2(r,z)\right)\,,\\
\rule{0pt}{1.5em}\omega_a(r,z) = \exp\left[-i (k\cdot \widehat x) \rho(r,z)\right]
\left[\left(\lambda_1(r,z) + i (k\cdot \widehat x) \eta_1(r,z)\right) \widehat x^a
- i \left((k\cdot \widehat x) \widehat x^a - k^a\right)\zeta(r,z)\right]\,.
\end{array}
\right.
\label{ansphi1}
\end{equation}
In the above equation, the 3-vector $k_a$ denotes the Skyrmion rotational velocity
$$
K=k_a\sigma^a/2=-i U^\dagger dU/dt\,,
$$
and the field $\rho$ is the same that appears in the ansatz for the gauge fields in eq.~(\ref{ansk1}).
The above ansatz can be obtained,
similarly to the one for the static solution, by imposing time-inversion, parity, and
cylindrical symmetry.

Plugging the ansatz in the 5d lagrangian one obtains the collective coordinates lagrangian
\be
L = -M +\frac{\lambda}2\,k_ak^a\,,
\label{eq:Lag}
\ee
where $M$ is the Skyrmion mass and $\lambda$ is its moment of inertia.
The latter receives a contribution from the gauge Lagrangian
\begin{eqnarray}
\displaystyle
\lambda_G &=& 16\pi M_5\frac13\int_0^\infty dr\int^{z_{\IR}}_{z_{\UV}}dz\,\left\{a(z)\left[
-\left(D_{\bar\mu}\rho\right)^2
-r^2\left(\partial_{\bar\mu} Q\right)^2
-2Q^2
-\frac{r^2}4B_{{\bar\mu}{\bar\nu}}B_{{\bar\mu}{\bar\nu}}\right.\right.\nn\\
&& \displaystyle
\left.\left.
+r^2\left(D_{\bar\mu}\chi\right)^2+
\frac{r^2}2\left(\partial_{\bar\mu}w\right)^2
+\left(\chi_{(x)}\chi_{(x)}+w^2\right)\left(1+\phi_{(x)}\phi_{(x)}\right)
-4w\phi_{(x)}\chi_{(x)}
\right]\right.\nn\\
&& \displaystyle
\left.
+\gamma L
\bigg[
-2\epsilon^{\bar\mu\bar\nu}D_{\bar\mu}\rho\,\chi_{(x)}\left(D_{\bar\nu}\phi\right)_{(x)}
+2\epsilon^{\bar\mu\bar\nu}\partial_{\bar\mu}\left(r\,Q\right)\,\chi_{(x)}\epsilon^{(xy)}\left(D_{\bar\nu}\phi\right)_{(y)}\right.\nn\\
&& \displaystyle
\left.
-w\left(\frac12\epsilon^{\bar\mu\bar\nu}B_{\bar\mu\bar\nu}\left(\phi_{(x)}\phi_{(x)}-1\right)
+r\,Q\epsilon^{\bar\mu\bar\nu}A_{\bar\mu\bar\nu}
\right)
+2r\,Q\epsilon^{\bar\mu\bar\nu} D_{\bar\mu}\rho\partial_{\bar\nu}\left(\frac{s}{r}\right)
\bigg]
\right\}\,,\label{eq:lambdagauge}
\end{eqnarray}
and a contribution from the scalar, which is given by
\begin{eqnarray}
\lambda_\Phi &=& \frac{16}{3}\pi M_5 \int_0^\infty dr \int_{z_{\UV}}^{z_{\IR}} dz \left\{
a^3(z)\left[-\frac{r^2}{4} \lambda^* \lambda (D_{\bar\mu}\rho)(D_{\bar\mu}\rho)
-\frac{r^2}{4} (D_{\bar\mu}\eta)^*(D_{\bar\mu}\eta)
-\frac{r^2}{2} (\partial_{\bar\mu}\zeta)(\partial_{\bar\mu}\zeta)\right.\right.\nn\\
&& + \frac{r^2}{4} D_{\bar\mu}\rho\left[
\lambda (D_{\bar \mu} \eta)^* - \eta (D_{\bar \mu} \lambda)^* + {\rm h.c.}\right]
- \frac{1}{2} \eta^* \eta + \frac{1}{8} (\eta \phi^* - \phi\eta^*)^2
-\frac{r^2}{8}(\chi \lambda^* - \lambda \chi^*)^2\nn\\
&& \left.\left.-i \zeta(\eta \phi^* - \phi \eta^*) - \frac{1}{2} \zeta^2(1+ \phi^* \phi)\right]
-r^2 a^5(z) M^2_{Bulk}\left[\frac{1}{4} \eta^* \eta + \frac{1}{2} \zeta^2\right]\right\}\,,
\end{eqnarray}
where we defined
\begin{equation}
 \eta = \eta_1 + i \eta_2\,,\qquad
\qquad D_{\bar \mu} \eta \equiv \partial_{\bar \mu} \eta - i A_{\bar \mu} \eta\,.
\end{equation}
while the other notations are defined in appendix~\ref{app:EOM}.

\subsection{Numerical Results}

The soliton solution can not be determined analytically, however, it can be studied
numerically by using the techniques described in~\cite{Panico:2008it}.
In the massless case it was found that, due to the peculiarity of the $5$d gauge action,
the soliton solution is stabilized thanks to the presence of the CS term \cite{Pomarol:2008aa}.
This peculiar feature disappears once we modify the action by the
introduction of the bulk scalar field and, in the present model, we checked in our numerical
analysis that the Skyrmion size is stable even if the CS term is not present.

From the soliton solution we can extract the electromagnetic and axial properties
of the nucleons, which are encoded in a set of form factors which parametrize the matrix
element of the currents on two nucleon states.

The chiral currents can be determined by computing the variation of the action
with respect to the sources ${\bf l}_\mu$ and ${\bf r}_\mu$. It is simple to show that
the action describing the scalar field $\Phi$ does not contribute to the currents,
which are exactly the same as in the massless pion case:
\begin{equation}
J_{L\, \mu}^{a}\,=\,M_5\big(a(z)L_{\mu\,5}^a\big)\left|_{z=z_{\UV}}\right.\ ,
\;\;\;\;\;{\widehat J}_{L\, \mu}\,=\,M_5\big(a(z){\widehat L}_{\mu\,5}\big)\left|_{z=z_{\UV}}\right.\, ,
\label{cur0}
\end{equation}
and analogously for $R$.

The isoscalar and isovector form factors are defined through the relations
\begin{eqnarray}
\displaystyle
\langle N_f({\vec q}/2) | J^0_{S}(0) | N_i(-{\vec q}/2)\rangle &=&  G_{E}^{S}({\vec q\,}^2) \chi_f^\dagger \chi_i\,,\nn\\
\displaystyle
\langle N_f({\vec q}/2) | J^i_{S}(0) | N_i(-{\vec q}/2)\rangle &=& i\, \frac{G_M^{S}({\vec q\,}^2)}{2 M_N} \chi_f^\dagger 2 ({\vec S} \times {\vec q})^i \chi_i\,,\nn\\
\displaystyle
\langle N_f({\vec q}/2) | J^{0 a}_{V}(0) | N_i(-{\vec q}/2)\rangle &=& G_{E}^{V}({\vec q\,}^2) \chi_f^\dagger \left(2I^a\right) \chi_i\,,\nn\\
\displaystyle
\langle N_f({\vec q}/2) | J^{i a}_{V}(0) | N_i(-{\vec q}/2)\rangle &=& i\, \frac{G_M^{V}({\vec q\,}^2) }{2 M_N}
\chi_f^\dagger 2 ({\vec S}  \times {\vec q})^i \left(2I^a\right) \chi_i\,,
\label{bf}
\end{eqnarray}
where the currents are given by $J_V^a = J_R^a + J_L^a$ and
$J_S = 1/3({\widehat J}_R + {\widehat J}_L)$, and
we used the notation $\chi_{i,f}$ for the nucleon spin/isospin
vectors of state (normalized to $\chi^\dagger \chi = 1$) and the definition
$({\vec S} \times {\vec q})^i \equiv \varepsilon^{ijk} S^j q^k$.
From the axial current $J_A^a = J_R^a - J_L^a$, we define the axial form factors
\begin{eqnarray}
\langle N_f({\vec q}/2) | J_A^{i,a}(0) | N_i(-{\vec q}/2)\rangle &=&
\chi_f^\dagger\! \left[ \frac{E}{M_N} G_A({\vec q\,}^2) S^i_T
+ \!\left(\!G_A({\vec q\,}^2) - \frac{{\vec q\,}^2}{4 M_N^2} G_p({\vec q\,}^2)\right)
\!\!S^i_L\right]\!\! I^a \chi_i\,,\hspace{1.5em}\\
\langle N_f({\vec q}/2) | J_A^{0,a}(0) | N_i(-{\vec q}/2)\rangle &=&0
\label{bfa}
\end{eqnarray}
where ${\vec S}_T \equiv {\vec S} - \hat{{\vec{q}}}\  {\vec S}\cdot{\hat{\vec{q}}}$ and
${\vec S}_L \equiv \hat{{\vec{q}}}\  {\vec S}\cdot{\hat{\vec{q}}}$
are the transverse and the longitudinal components of the spin operator.

To find the explicit expressions for the form factors we need to plug the ansatze for the
soliton solution into the definitions of the currents (\ref{cur0}) and then perform the
quantization of the soliton solution as explained in \cite{Panico:2008it}. The result
is the same as in the massless pion case:
\begin{eqnarray}
\displaystyle
&&G_E^S\,=\,-\frac{N_c}{6\pi\gamma L}\int dr\,r\,j_0(qr)\left(a(z)\partial_zs\right)_{UV}\nn\\
&&G_E^V\,=\,\frac{4\pi M_5}{3\lambda}\int dr\,r^2\,j_0(qr)\left[a(z)\left(\partial_z w-2\left(D_z\chi\right)_{(2)}\right)\right]_{UV}\nn\\
&&G_M^S\,=\,\frac{8\pi M_NM_5}{3\lambda}\int dr\,r^3\,\frac{j_1(qr)}{qr}\left(a(z)\partial_zQ\right)_{UV}\nn\\
&&G_M^V\,=\,\frac{M_N\,N_c}{3\pi L\gamma}\int dr\,r^2\,\frac{j_1(qr)}{qr}\left(a(z)\left(D_z\phi\right)_{(2)}\right)_{UV}\nn\\
&&G_A\,=\,\frac{M_N}{E}
\frac{N_c}{3\pi\gamma L}
\int dr\,r\left[
a(z)\frac{j_1(qr)}{qr}
\left(\left(D_z\phi\right)_{(1)}-r\,A_{zr}\right)
-a(z)\left(D_z\phi\right)_{(1)} j_0(qr)
\right]_{UV}
\label{cff}
\end{eqnarray}
where $j_n$ are spherical Bessel functions.

\begin{table}
\centering
\begin{tabular}{l@{\hspace{.275em}}l@{\hspace{.275em}}l@{\hspace{.275em}}c}
\hline
& Experiment & AdS$_5$ & Deviat.\\
\hline
$M_N$ & $940\ {\rm MeV}$ & $1090\ {\rm MeV}$ & $16\%$\\
$\mu_S$ & $0.44$ & $0.43$ & $2\%$\\
$\mu_V$ & $2.35$ & $1.18$ & $100\%$\\
$g_A$ & $1.25$ & $0.58$ & $100\%$\\
$\sqrt{\langle r^2_{E,S}\rangle}$ & $0.79\ {\rm fm}$ & $0.82\ {\rm fm}$ & $4\%$\\
$\sqrt{\langle r^2_{E,V}\rangle}$ & $0.93\ {\rm fm}$ & $0.97\ {\rm fm}$ & $4\%$\\
$\sqrt{\langle r^2_{M,S}\rangle}$ & $0.82\ {\rm fm}$ & $0.84\ {\rm fm}$ & $2\%$\\
$\sqrt{\langle r^2_{M,V}\rangle}$ & $0.87\ {\rm fm}$ & $0.87\ {\rm fm}$ & $0.5\%$\\
$\sqrt{\langle r^2_{A}\rangle}$ & $0.68\ {\rm fm}$ & $0.65\ {\rm fm}$ & $5\%$\\
\hline
\end{tabular}
\hspace{1.5em}
\begin{tabular}{l@{\hspace{.275em}}l@{\hspace{.275em}}l@{\hspace{.275em}}c}
\hline
& Experiment & AdS$_5$ & Deviat.\\
\hline
$M_N$ & $940\ {\rm MeV}$ & $1104\ {\rm MeV}$ & $17\%$\\
$\mu_S$ & $0.44$ & $0.43$ & $3\%$\\
$\mu_V$ & $2.35$ & $1.15$ & $100\%$\\
$g_A$ & $1.25$ & $0.59$ & $100\%$\\
$\sqrt{\langle r^2_{E,S}\rangle}$ & $0.79\ {\rm fm}$ & $0.84\ {\rm fm}$ & $6\%$\\
$\sqrt{\langle r^2_{E,V}\rangle}$ & $0.93\ {\rm fm}$ & $1.02\ {\rm fm}$ & $9\%$\\
$\sqrt{\langle r^2_{M,S}\rangle}$ & $0.82\ {\rm fm}$ & $0.86\ {\rm fm}$ & $5\%$\\
$\sqrt{\langle r^2_{M,V}\rangle}$ & $0.87\ {\rm fm}$ & $0.86\ {\rm fm}$ & $1\%$\\
$\sqrt{\langle r^2_{A}\rangle}$ & $0.68\ {\rm fm}$ & $0.68\ {\rm fm}$ & $0.2\%$\\
\hline
\end{tabular}
\caption{Prediction for the static nucleon observables with the parameter values
fixed by the fit on the mesonic observables
(RMSE fit for the left table and `maximal deviation' fit for the
right table).}\label{baryon results}
\end{table}
By employing suitable numerical techniques,
the 2d EOM \footnote{The EOM for the 2d fields are reported in appendix~\ref{app:EOM}.}
obtained by varying the soliton mass $M$ and its moment of inertia $\lambda$
can be solved, and both the static and slowly-rotating Skyrmion solution computed.
The numerical predictions for the static nucleon observables are listed in
table~\ref{baryon results}. In the analysis we used the values of the microscopic parameters
obtained from the fits on the mesonic observables presented in section~\ref{sec:fit}.

The numerical results for the two sets of microscopic parameters considered show
very similar deviation patterns from the data.
Many of the numerical predictions are very close to the experimental results, although
the magnetic vector moment $\mu_V$ and the axial coupling $g_A$ present a deviation
of order $100\%$. We notice, however, that the overall agreement with the data
(root mean square error $~45\%$) is still compatible with the possibility of having
sizable $1/N_c$ corrections, which can not be excluded given that
$N_c=3$. \footnote{It is interesting to notice that using a different approach to the quantization
of the collective coordinates, as suggested in \cite{Amado:1986ef},
one gets much better predictions for
$\mu_V$ and $g_A$. With this procedure, the predictions for $\mu_V$ and $g_A$ are
rescaled by a factor $5/3$, thus agreeing with the data at the $20\%$ level,
while all the other observables are unchanged. We have no reason to believe that
the modified quantization procedure correctly captures the $1/N_c$ corrections,
nevertheless, this result seems to point out that large corrections
could indeed be responsible for the deviations of $\mu_V$ and $g_A$.}
By comparing the present results with the ones found in the simplified model
without a pion mass~\cite{Panico:2008it}, we see that all the observables
show an improved agreement with the data except for the $\mu_V$ and $g_A$,
whose deviations become significantly larger.

Form the qualitative point of view, we remark that an important
check of the validity of the description of
baryons as solitons is the behavior of the electric and magnetic vector radii,
namely $r_{E,V}$ and $r_{M,V}$. These two quantities are expected to be divergent in the chiral
limit, as explicitly verified in~\cite{Panico:2008it}, while they should become finite
once a pion mass is introduced, as we find in the present model.

\section{Conclusions}\label{sec:Conclusions}

We have shown that it is rather easy to construct a model of holographic
QCD which describes at the same time the pion mass, the QCD anomalies and the baryons
as topological solitons. After introducing an explicit minimal model we have studied
its phenomenology in both the mesonic and baryonic sector and found a significant
level of agreement. In extreme synthesis, our
result is that the general picture on the holographic QCD models outlined in the
Introduction survives unchanged to the inclusion of the pion mass.

Few unexpected results have been found, however, that is worth discussing.
In Sect.~\ref{sec:MassivePions} we saw that our model easily reproduces the mass of the $\pi(1300)$ meson,
a task that was impossible to achieve in the original scenario \cite{DaRold:2005vr}.
This came because of the change in the IR boundary conditions and gives a phenomenological
support to this modification, that we had instead motivated on purely theoretical grounds.
It is also remarkable that the other predictions are almost unaffected so that
all the valid phenomenology of the original construction is retained. The ``new''
observables that were absent in the original model, {\it{i.e.}} the anomalous
parity couplings originating from the CS term, also show a good agreement with the
observations. Our results in the baryonic sector, shown in table~\ref{baryon results},
are also surprising,
especially if compared with the ones obtained in the chiral-symmetric
case \cite{Pomarol:2007kr,Pomarol:2008aa,Panico:2008it}.
For all the observables except $\mu_V$ and $g_A$, a significant improvement is found in the
agreement with observations. The isovector radii, that have became finite due to the presence
of the pion mass, are also extremely well predicted. The situation has got significantly worst,
on the contrary, for $\mu_V$ and $g_A$ that have became a factor of $2$ smaller than the
observations. \footnote{Actually, the discrepancy is almost exactly given by a factor of $2$. By looking
at table~\ref{baryon results} one could imagine a factor $2$ mistake in the problematic predictions, we are
however confident of our calculations.}
This failure persists in both the best fit points that we have used as input
parameters in table~\ref{baryon results}, so that it is probably a robust feature.
It might signal that the model is incomplete, but it might also be attributed to anomalously large $1/N_c$ corrections.

Some final comments on the theoretical implications of our results. The microscopic
origin of holographic QCD models is basically unknown, even though the holographic implementation of the chiral
symmetry provides a robust (but purely technical) connection with AdS/CFT. The success
of the Regge phenomenology suggests, independently of AdS/CFT, the dual of large-$N_c$ QCD being a
string model and the validity of the holographic QCD approach suggests that this string
model should contain a sector that is well described by a 5d field theory similar to the one we
have considered. In the case of exact chiral symmetry, the Sakai--Sugimoto model 
\cite{Sakai:2004cn,hs} provides a partial realization of this idea because it is 
equivalent to holographic QCD for what the physics of the light meson is
concerned. \footnote{See \cite{Becciolini:2009fu} for a precise justification of this equivalence.} 
It is on the contrary different, and problematic, in the baryonic sector \cite{Pomarol:2007kr,Pomarol:2008aa,Panico:2008it,Cherman:2009gb} and its phenomenology 
in the sector of higher spin states (where it genuinely
differs from holographic QCD and shows its stringy nature) seems not very promising \cite{hs}.
The inclusion of the explicit chiral breaking considered in the present paper provides an additional
piece of information. The chiral symmetry is not explicitly broken in Sakai--Sugimoto, and even 
 though it was possible
to include the explicit breaking by some deformation, the resulting model could not reduce to a field-theoretical
model such as the one we have considered. The reason for this is that the Left and Right global groups are
localized, in Sakai--Sugimoto, at two different boundaries of the 5d space and the quark mass
spurion $M$ is unavoidably a non local object which is impossible to describe in a field theoretical language.
Therefore, any string model that incorporated ours, inheriting its phenomenology, would be deeply
different from Sakai--Sugimoto; it might be worth trying to construct one following a bottom--up approach.

\section*{Acknowledgments}

We would like to thank R.~Rattazzi for useful discussions and
A.~Pomarol for his collaboration in the early stage of the project.
O.~D. is grateful to EPFL for hospitality during the completion of part of this work.

\appendix

\section{The Effective Action for the Pion}\label{app:HoloAction}

In this appendix we will compute the effective action for the pion at ${\cal O}(p^4)$.
Given that a complete treatment of the $U(1)_A$ anomaly is not included in the model,
in the following we will neglect the Goldstone boson related to this symmetry, namely
the $\eta'$ field, and we will only consider a model with a chiral symmetry
$SU(2)_L \times SU(2)_R$.

An efficient way to perform the computation is to use the holographic
approach presented in~\cite{Panico:2007qd}. At tree-level, the holographic action
for the pion is given by the 5d action for the gauge and the scalar fields
(eqs.~(\ref{Sg}) and (\ref{sscalar})), where the 5d fields satisfy the bulk EOM's
with the usual IR boundary conditions given in eqs.~(\ref{IRgaugebc})
and (\ref{irboundary condition}). The UV conditions are modified as
\begin{equation}
\left\{
\begin{array}{l}
\left. L_\mu\right|_{z_{\UV}} = U(x)(l_\mu + i \partial_\mu) U(x)^\dagger\\
\left. R_\mu\right|_{z_{\UV}} = r_\mu\\
\left. \Phi\right|_{z_{\UV}} = U(x) \left(\frac{z_{\UV}}{z_{\IR}}\right)^{\Delta_-}M_q \I
\end{array}
\right.\,,\label{eq:UVbcholo}
\end{equation}
where $U(x)$ represents the 4d Goldstone matrix, which transforms as
\begin{equation}
U(x) \rightarrow g_R U(x) g_L^\dagger
\end{equation}
under a chiral $SU(2)_L \times SU(2)_R$ 4d transformation. Notice that we are not
interested in possible terms involving the scalar and pseudoscalar sources,
so we did not include any source term in the UV condition for
the scalar field in eq.~(\ref{eq:UVbcholo}).

To derive the complete effective action for the pion one would need to solve
the full bulk EOM for the 5d fields. However, due to the presence of interaction terms,
this can be done only perturbatively. As usually done in ${\bf \chi PT}$, we use an expansion
in powers of the momentum $p$ and we treat the external sources $l_\mu$ and $r_\mu$
as ${\cal O}(p)$ terms, while $M_q$ will be treated as an ${\cal O}(p^2)$ term.

We expand the solutions of the EOM using a mixed momentum--space representation
\begin{equation}
\left\{
\begin{array}{l}
V_\mu = f_V^0(z) \widehat V_\mu(p) + V^{(3)}_\mu(p, z, \widehat V, \widehat A, U, M_q)\,,\\
\rule{0pt}{1.25em}A_\mu = f_A^0(z) \widehat A_\mu(p) + A^{(3)}_\mu(p, z, \widehat V, \widehat A, U, M_q)\,,\\
\rule{0pt}{1.25em}\Phi = \Phi^{(0)}(z) + \Phi^{(2)}(p, z, \widehat V, \widehat A, U, M_q)
+ \Phi^{(4)}(p, z, \widehat V, \widehat A, U, M_q)\,,
\end{array}
\right.
\end{equation}
where we denoted by $\widehat V$ and $\widehat A$ the values of the vector and axial gauge
fields at the UV boundary. Notice that, due to the tensorial structure, the gauge field
solutions start at ${\cal O}(p)$ and their next to leading terms are of ${\cal O}(p^3)$,
while the scalar field can be expanded in terms of ${\cal O}(p^{2n})$.
All the terms in the expansion of the gauge fields satisfy the same IR boundary
conditions as the original fields:
\begin{equation}
\left\{
\begin{array}{l}
\partial_z V_\mu(x, z_{\IR}) = 0\,,\\
\rule{0pt}{1.25em}A_\mu(x, z_{\IR}) = 0\,.
\end{array}
\right.
\end{equation}
Instead, for the scalar field we have
\begin{equation}
\left\{
\begin{array}{l}
\Phi^{(0)}(x, z_{\IR}) = \xi \I\,,\\
\rule{0pt}{1.25em}\Phi^{(i)}(x, z_{\IR}) = 0 \qquad {\rm for\ } i \geq 2\,.
\end{array}
\right.
\end{equation}
The UV boundary conditions for the gauge fields are chosen so that the
higher terms in the expansion vanish, while for the leading terms we have
\begin{equation}
f_{V,A}^0(z_{\UV}) = 1\,.
\end{equation}
For the scalar field we impose
\begin{equation}
\left\{
\begin{array}{l}
\Phi^{(0)}(z_{\UV}) = 0\,,\\
\rule{0pt}{1.25em}\Phi^{(2)}(z_{\UV}) =
\displaystyle U \left(\frac{z_{\UV}}{z_{\IR}}\right)^{\Delta_-} M_q \I\,,\\
\rule{0pt}{1.25em}\Phi^{(4)}(z_{\UV}) = 0 \qquad {\rm for\ } i \geq 4\,.
\end{array}
\right.
\end{equation}
By using the bulk EOM's and the boundary conditions for the fields, one can verify that
the terms of ${\cal O}(p^3)$ in the expansion for the gauge fields and the ones of order
${\cal O}(p^4)$ in the expansion for the scalar field do not contribute to the
effective action for the pion at ${\cal O}(p^4)$ (see the discussion in~\cite{Panico:2007qd}).

As a first step of the derivation of the effective action,
we will consider the contributions coming
from the 5d gauge action in eq.~(\ref{Sg}). It is convenient to rewrite eq.~(\ref{Sg})
in the form
\begin{eqnarray}
S_g &=& -2 M_5 \int_{UV} d^4 x\ a(z_{\UV})\, {\rm Tr}\left[V_\mu \partial_z V^\mu
+ A_\mu \partial_z A^\mu\right]\nn\\
&& - M_5 \int d^4 x\ \int dz a(z)\, {\rm Tr}\left[L_{\mu\nu} L^{\mu\nu}
+ R_{\mu\nu} R^{\mu\nu}\right]\,.\label{Sgholo}
\end{eqnarray}
The leading terms in the expansion of the 5d fields satisfy the equations
\begin{equation}
\left\{
\begin{array}{l}
\partial_z\left(a(z) \partial_z f^0_V(z)\right) = 0\,,\\
\rule{0pt}{1.25em}\partial_z\left(a(z) \partial_z f^0_A(z)\right) - 2 a^3(z) (\Phi^{(0)})^2 f^0_A(z) = 0\,,\\
\rule{0pt}{1.25em}\partial_z \left(a^3(z) \partial_z \Phi^{(0)}(z)\right) - a^5(z) M_\phi^2 \Phi^{(0)}(z) = 0\,.
\end{array}
\right.
\end{equation}
The solution for the $\Phi^{(0)}$ field can be obtained from eqs.~(\ref{PhiVEV})
and (\ref{Phicoeff}) by setting $M_q = 0$.
The equation for the vector gauge field admits the simple solution $f_V^0(z) = 1$,
while the equation for $f_A^0$ in general can not be solved analytically.

By substituting the above expressions in the gauge action and
using the relation
\begin{equation}
\widehat A_\mu = \frac{i}{2} U \left(D_\mu U\right)^\dagger
= -\frac{i}{2} \left(D_\mu U\right) U^\dagger
\equiv u_\mu\,,
\end{equation}
we find that the first line in eq.~(\ref{Sgholo}) gives
the kinetic term for the pion, from which we can extract the pion decay constant
\begin{equation}
f_\pi^2 = \left. -2 M_5 a(z) \partial_z f_A^0(z)\right|_{z=z_{\UV}}\,.
\end{equation}
From the second line in eq.~(\ref{Sgholo}) we get contributions to the
${\cal O}(p^4)$ terms in the pion effective Lagrangian.
In the standard form of ${\bf \chi PT}$~\cite{Gasser:1984gg}
we get the following contributions
\begin{equation}
\left\{
\begin{array}{l}
L_1^{(g)} = \displaystyle\frac{M_5}{16}\int dz\, a(z)\left(1 - (f_A^0)^2\right)^2\\
\rule{0pt}{1.5em}L_{10}^{(g)} = \displaystyle-\frac{M_5}{2}\int dz\, a(z)\left(1 - (f_A^0)^2\right)
\end{array}
\right.
\,, \qquad
\left\{
\begin{array}{l}
L_2^{(g)} = 2 L_1^{(g)}\\
L_3^{(g)} = -6 L_1^{(g)}\\
L_9^{(g)} = - L_{10}^{(g)}
\end{array}
\right.\,.
\end{equation}

Now we consider the contributions to the effective action coming from the
scalar action in eq.~(\ref{sscalar}). To derive the action we need to compute
the $\Phi^{(2)}$ term in the expansion of the 5d scalar field. This term satisfies
the bulk EOM
\begin{equation}\label{eq:phi2}
-\frac{1}{a^3(z)} \partial_z \left(a^3(z) \partial_z \Phi^{(2)}\right)
+ a^2(z) M_\phi^2 \Phi^{(2)} = - 2 \Phi^{(0)} \left[-2 (f_A^0)^2 u_\mu u^\mu
-i f_A^0 D_\mu u^\mu\right]\,,
\end{equation}
where we defined $D_\mu u^\mu \equiv \partial_\mu u^\mu - i [\widehat V_\mu, u^\mu]$.
The solution can be split into two parts:
\begin{equation}
\Phi^{(2)} = \Phi^{(2)}_M + \Phi^{(2)}_0\,,
\end{equation}
where $\Phi^{(2)}_M$ is a solution of the homogeneous part of eq.~(\ref{eq:phi2})
with boundary conditions
\begin{equation}
\Phi^{(2)}_M(z_{\UV}) =
\displaystyle U \left(\frac{z_{\UV}}{z_{\IR}}\right)^{\Delta_-} M_q \I
\,, \qquad \Phi^{(2)}_M(z_{\IR}) = 0\,,
\end{equation}
while $\Phi^{(2)}_0$ satisfies eq.~(\ref{eq:phi2}) with boundary conditions
\begin{equation}
\Phi^{(2)}_0(z_{\UV}) = \Phi^{(2)}_0(z_{\IR}) = 0\,.
\end{equation}
The solution for $\Phi^{(0)}_M$ is simply given by eqs.~(\ref{PhiVEV})
and (\ref{Phicoeff}) with the choice $\xi = 0$. The solution for
$\Phi^{(0)}_0$ can not be found analytically, and we will parametrize it as
\begin{equation}
\Phi^{(0)}_0 = f_R^{(2)}(z) u_\mu u^\mu + f_I^{(2)}(z) D_\mu u^\mu\,.
\end{equation}

From the action for the scalar field we get an ${\cal O}(p^2)$ contribution to
the pion effective action, which can be written as
\begin{equation}\label{eq:scalarorder2}
S_\phi^{(2)} = M_5 \int_{UV} d^4 x\, a^3(z) {\rm Tr}\left[
{\Phi^{(2)}_M}^\dagger \partial_z \Phi^{(0)} + {\rm h.c.}\right]\,.
\end{equation}
To obtain this expression we integrated by parts the terms containing derivatives
with respect to $z$ and we used the bulk EOM and the boundary conditions for the
terms in the scalar field expansion.
Eq.~(\ref{eq:scalarorder2}) corresponds to a mass term for the pion, which
in the limit $z_{\UV} \rightarrow 0$ becomes
\begin{equation}\label{eq:piMassTermLag}
S_\phi^{(2)} = 2 \frac{M_5}{L}\, \alpha\, \xi \int d^4x\, {\rm Tr}\left[
U M_q^\dagger + M_q U^\dagger\right]\,.
\end{equation}
The expression for the pion mass and for the pion decay constant can be easily
computed as an expansion in the parameter $\xi$. Some approximate expressions are
reported in eq.~(\ref{eq:fpi_mpi}).

Computing the ${\cal O}(p^4)$ terms in the action we get
\begin{eqnarray}
S_\phi^{(4)} & = & M_5 \int d^4x \int dz\, a^3(z) \Big\{
2 {\rm Tr}\left[\Phi^{(0)} f_I^{(2)}(z) f_A^0(z) D_\mu u^\mu D_\nu u^\nu\right]\nn\\
&& \hspace{9.5em} + 4 {\rm Tr}\left[\Phi^{(0)} f_R^{(2)}(z) (f_A^0(z))^2 u_\mu u^\mu u_\nu u^\nu\right]\nn\\
&& \hspace{9.5em} + i {\rm Tr}\left[\Phi^{(0)} \left({\Phi^{(2)}_M}^\dagger - {\Phi^{(2)}_M}\right)
f_A^0(z) D_\mu u^\mu\right]\nn\\
&& \hspace{9.5em} + 2 {\rm Tr}\left[\Phi^{(0)} \left({\Phi^{(2)}_M}^\dagger + {\Phi^{(2)}_M}\right)
(f_A^0(z))^2 u_\mu u^\mu\right]\Big\}\nn\\
&& + \frac{M_5}{2}\int_{UV} d^4x\, a^3(z) {\rm Tr}
\left[ \partial_z(f_R^{(2)}(z) u_\mu u^\mu - i f_I^{(2)}(z) D_\mu u^\mu)\Phi^{(2)}_M
+ {\rm h.c.} \right]\,.
\end{eqnarray}
The ${\cal O}(p^4)$ action for the pion can be simplified by using the
EOM for the pion field coming from the ${\cal O}(p^2)$ effective action.
Using the standard notation of ${\bf \chi PT}$~\cite{Gasser:1984gg},
the kinetic and mass terms for the pion are written as
\begin{equation}\label{eq:piO2_chiPT}
{\cal L} = \frac{f_\pi^2}{4} \left\{
{\rm Tr}\left[(D_\mu U)^\dagger D^\mu U\right]
+ {\rm Tr}\left(\chi^\dagger U + \chi U^\dagger\right)\right\}\,.
\end{equation}
From this Lagrangian we get the EOM for $U$:
\begin{equation}
(D_\mu D^\mu U) U^\dagger - U (D_\mu D^\mu U)^\dagger + U \chi^\dagger - \chi U^\dagger = 0\,,
\end{equation}
which can be rewritten as
\begin{equation}
4 i D_\mu u^\mu = \chi U^\dagger - U \chi^\dagger\,.
\end{equation}
Moreover, by comparing eq.~(\ref{eq:piO2_chiPT}) and eq.~(\ref{eq:piMassTermLag}), we can extract
the relation between $M_q$ and $\chi$, which, in the $z_{\UV} \rightarrow 0$, limit reads
\begin{equation}
M_q = \frac{L}{M_5 \alpha \xi} \frac{f_\pi^2}{8} \chi\,.
\end{equation}

By using these relations we get the scalar contributions to the coefficients
of the ${\cal O}(p^4)$ effective pion action
\begin{equation}
\left\{
\begin{array}{l}
L_3^{(\phi)} = \displaystyle\frac{M_5}{8} \int dz\ a^3(z) {\rm Tr}\left(\Phi^{(0)}\right) f_R^{(2)}(z) (f_a^0(z))^2\,,\\
L_5^{(\phi)} = \displaystyle\frac{\kappa M_5}{8}\left\{
\left.a^3(z) f_M^{(2)}(z) \partial_z f_R^{(2)}(z)\right|_{z_{\UV}}
\!\! + 2\!\! \int dz\ a^3(z) f_M^{(2)}(z) (f_a^0(z))^2 {\rm Tr}\left(\Phi^{(0)}\right)\right\}\,,\\
L_8^{(\phi)} = \displaystyle\frac{M_5}{16}\left\{
\!\left.2 \kappa a^3(z) f_M^{(2)}(z) \partial_z f_I^{(2)}(z)\right|_{z_{\UV}}
\!\! -\!\! \int dz\ a^3(z) f_a^0(z) (f_I^{(2)}(z) - 2 \kappa f_M^{(2)}(z))
{\rm Tr}\left(\Phi^{(0)}\right)\right\}\,,
\end{array}
\right.
\end{equation}
where $\kappa$ is defined by the relation $M_q = \kappa \chi$.

\section{The Equations of Motion}\label{app:EOM}

In this appendix we report the EOM for the 2d fields which appear in the ansatz
for the zero-mode soliton fluctuations in eqs.~(\ref{sts}), (\ref{ansk1})
and (\ref{ansphi1}), and we summarize the notation used in the paper.

\subsection{The Equations of Motion}

Before writing the EOM's for the 2d fields, it is useful to recall the residual
symmetries which survive after we choose the ansatze for the Skyrmion solution.
As already discussed in sec.~\ref{sec:staticsolution}, the ansatz preserves a
$U(1)$ local symmetry with 2d gauge field $A_\mu$, which corresponds to the 5d
gauge transformations given in eq.~(\ref{eq:resU1SU2}). The fields $\phi$,
$\lambda$, $\chi$ and $\eta$ are charged under this symmetry, thus it is convenient to
write the action and the EOM in terms of their covariant derivatives
\begin{equation}
\begin{array}{l@{\qquad}l}
D_\mu \phi = \partial_\mu \phi - i A_\mu \phi\,,
& D_\mu \chi = \partial_\mu \chi - i A_\mu \chi\,,\\
D_\mu \lambda = \partial_\mu \lambda - i A_\mu \lambda\,,
& D_\mu \eta = \partial_\mu \eta - i A_\mu \eta\,.
\end{array}
\end{equation}
There is also a second residual $U(1)$ generated by the $U(1)_{L,R}$ 5d transformations of
the form ${\widehat g}_R={\widehat g}$ and ${\widehat g}_L={\widehat g}^\dagger$ with
\be
{\widehat g} = \exp\left[i \beta(r,z)(k\cdot\x)\right]\,,
\label{eq:resU1U1}
\ee
whose associated gauge boson is $B_{\bar\mu}$ (see \cite{Panico:2008it}).
The $\rho$ field transforms as a Goldstone boson under this symmetry, so we can define
its covariant derivative as
\begin{equation}
D_\mu \rho = \partial_\mu \rho - B_\mu\,.
\end{equation}

The EOM's for the 2d fields can be easily found by substituting the ansatz for the
Skyrmion solution into the 5d action.
The EOM's for the fields which appear in the static soliton case
\be
\left\{
\begin{array}{l}
\displaystyle
\rule{0pt}{1.5em}D^{\bar\mu}\left(a(z) D_{\bar\mu} \phi\right)
+ \frac{a(z)}{r^2}\phi(1-|\phi|^2) + \frac{a^3(z)}{4} \lambda
(\lambda \phi^\dagger - \phi \lambda^\dagger) + i\gamma L \epsilon^{\bar\mu \bar\nu}
\partial_{\bar\mu}\left(\frac{s}{r}\right)D_{\bar\nu}\phi = 0\\
\displaystyle
\rule{0pt}{1.5em}\partial^{\bar\mu}\left(r^2 a(z) A_{\bar\mu\bar\nu}\right)
- a(z) \left(i \phi^\dagger D_{\bar\nu} \phi + h.c.\right)
- \frac{i}{4} a^3(z) r^2\left[\lambda^\dagger (D_{\bar\mu}\lambda) - (D_{\bar\mu}\lambda)^\dagger \lambda\right]\\
\rule{0pt}{1.25em}\hspace{3em}+ \gamma L \epsilon^{\bar\mu \bar\nu} \partial_{\bar\mu}\left(\frac{s}{r}\right)
(|\phi|^2-1)=0\\
\displaystyle
\rule{0pt}{1.5em}\partial_{\bar\mu}\left(a(z)\partial^{\bar\mu}s\right)
-\frac{\gamma L}{2 r}\epsilon^{\bar\mu\bar\nu}\left[
\partial_{\bar\mu}(-i \phi^\dagger D_{\bar\nu}\phi + h.c.) + A_{\bar\mu\bar\nu}\right]
= 0\\
\rule{0pt}{1.5em}D^{\bar\mu}\left(r^2 a^3(z) D_{\bar\mu} \lambda\right) - a^3(z) \phi(\lambda \phi^\dagger
- \phi \lambda^\dagger) -a^5(z) r^2 M^2_{Bulk} \lambda = 0
\end{array}
\right.\,,
\ee
while the equations for the fields which are turned on for the rotating Skyrmion are
\be
\left\{
\begin{array}{l}
\displaystyle
\partial^{\bar\mu}(r^2 a(z) \partial_{\bar\mu} w) - 2 a(z)\left[
w(1+\left|\phi\right|^2) - \chi\phi^\dagger-\phi\chi^\dagger\right]
+\gamma L \epsilon^{\bar\mu \bar\nu}\!\left[
\frac{1}{2}(\left|\phi\right|^2 - 1)
B_{\bar\mu \bar\nu}+ r Q
A_{\bar\mu \bar\nu}\right]=0\\
\displaystyle
\rule{0pt}{1.5em}D^{\bar\mu}(r^2 a(z) D_{\bar\mu} \chi) + a(z)\left[2w\phi
-(1+\left|\phi\right|^2)\chi\right]
+ \frac{1}{4} a^3(z) r^2 \lambda(\lambda \chi^\dagger - \chi \lambda^\dagger)\\
\rule{0pt}{1.25em}\hspace{3em}-\gamma L \epsilon^{\bar\mu \bar\nu}(D_{\bar\mu}\phi)
\left[i \partial_{\bar\nu}(r Q)+ D_{\bar\nu}\rho\right]=0\\
\displaystyle
\rule{0pt}{1.5em}\frac{1}{r}\partial^{\bar\mu}(r^2 a(z) \partial_{\bar\mu} Q)
-\frac{2}{r}a(z) Q\\
\displaystyle
\hspace{3em}-\frac{\gamma L}{2}\epsilon^{\bar\mu \bar\nu}\Big[
(i D_{\bar\mu}\phi (D_{\bar\nu}\chi)^\dagger + h.c.)
+ \frac{1}{2}A_{\bar\mu \bar\nu}(2 w - \chi\phi^\dagger-\phi\chi^\dagger)
- \frac{2}{\alpha^2}D_{\bar\mu}\rho\, \partial_{\bar\nu}\left(\frac{s}{r}\right)
\Big]=0\\
\displaystyle
\rule{0pt}{1.5em}\partial^{\bar \mu}(a(z) D_{\bar\mu}\rho)
- \frac{1}{4} a^3(z) \left[(\eta \lambda^\dagger + \lambda \eta^\dagger)
-2 i \zeta(\phi\lambda^\dagger - \lambda \phi^\dagger)\right]\\
\displaystyle
\rule{0pt}{1.25em}\hspace{3em}-\frac{\gamma L}{2}\epsilon^{\bar\mu \bar\nu}\Big[
\left(D_{\bar\mu}\phi(D_{\bar\nu}\chi)^\dagger + h.c.\right)
+\frac{i}{2}A_{\bar\mu \bar\nu}(\phi\chi^\dagger - \chi \phi^\dagger)
+\frac{2}{\alpha^2} \partial_{\bar\mu}(r Q)\partial_{\bar\nu}\left(\frac{s}{r}\right)\Big]=0\\
\displaystyle
\rule{0pt}{1.5em}\partial^{\bar\nu}\left(r^2 a(z)B_{\bar\nu\bar\mu}\right)
+ 2 a(z) D_{\bar\mu} \rho
+ \frac{1}{4} r^2 a^3(z)\left[2 \lambda^\dagger \lambda D_{\bar\mu}\rho
+ \left(\eta (D_{\bar\mu}\lambda)^\dagger - \lambda (D_{\bar\mu}\eta)^\dagger + {\rm h.c.}\right)\right]\\
\displaystyle
\hspace{3em}+\gamma L \epsilon^{\bar\mu\bar\nu}
\Big\{\left[(\chi-w \phi)(D_{\bar\nu}\phi)^\dagger + h.c.\right]
+(1-|\phi|^2)\partial_{\bar\nu} w -\frac{2 r}{\alpha^2} Q\, \partial_{\bar\nu}\left(\frac{s}{r}\right)
\Big\} =0\\
\rule{0pt}{1.5em}D^{\bar\mu}(a^3(z)r^2 D_{\bar\mu} \eta) - \lambda \partial^{\bar\mu}(a^3(z)r^2 D_{\bar\mu}\rho)
-2 a^3(z) r^2 (D^{\bar\mu}\rho)(D_{\bar\mu}\lambda)\\
\rule{0pt}{1.25em}\hspace{3em}-a^3(z) \eta(2 + |\phi|^2) + a^3(z) \phi(\phi \eta^\dagger + 4 i \zeta) - r^2 a^5(z) M^2_{Bulk} \eta =0\\
\rule{0pt}{1.5em}\partial^{\bar\mu}(r^2 a^3(z) \partial_{\bar\mu} \zeta)
-a^3(z)\left[\zeta(1+|\phi|^2)-i(\phi\eta^\dagger - \eta\phi^\dagger)\right]
-r^2 a^5(z) M^2_{Bulk} \zeta = 0\,.
\end{array}
\right.
\ee

In order to find suitable equations for the numerical analysis of the
solutions, the EOM's must be rewritten as a system of elliptic partial differential equations.
For this purpose we need to choose a gauge fixing condition for the
residual 2d $U(1)$ gauge symmetries. A possible choice is
a Lorentz gauge condition
\be
\partial^{\bar\mu} A_{\bar\mu}=0\,,
\qquad \quad
\partial^{\bar\mu} B_{\bar\mu}=0\,.
\label{eq:2Dgauge}
\ee
With this condition the equations for $A_{\bar\nu}$ become
$J^{\bar\nu} = \partial_{\bar\mu}\left(r^2 a A^{\bar\mu\bar\nu}\right)
= r^2 a \partial_{\bar\mu}\partial^{\bar\mu} A^{\bar\nu}
+\partial_{\bar\mu}(r^2 a) A^{\bar\mu\bar\nu}$ which is an elliptic equation
and a similar result is obtained for $B_{\bar\mu}$.

\subsection{The Boundary Conditions}

The derivation of the boundary conditions for the 2d fields
has been discussed in section~\ref{sec:staticsolution}.
Here we report the list of conditions we need to impose on the
scalar field components as well as the conditions for the
gauge fields, which are analogous to the ones for the massless
case~\cite{Panico:2008it}.

The IR and UV boundary conditions on the 2d fields follow from
the boundary conditions for the 5d fields
(eqs.~(\ref{IRgaugebc}) and (\ref{UVgaugebc}) with vanishing sources
for the gauge fields,
eqs.~(\ref{irboundary condition}) and (\ref{uvboundary condition}) for the scalar)
and from the gauge choice in eq.~(\ref{eq:2Dgauge}).
They are given explicitly by
\be
z=z_{\IR}\ :
\quad
\left\{
\begin{array}{l}
\phi_1 = 0\\
\partial_2 \phi_2 = 0\\
A_1 = 0\\
\partial_2 A_2 = 0\\
\partial_2 s = 0\\
\lambda = -2 i \xi
\end{array}
\right.
\qquad\qquad
\left\{
\begin{array}{l}
\chi_1 = 0\\
\partial_2 \chi_2 = 0\\
\partial_2 w = 0\\
\partial_2 Q = 0
\end{array}
\right.
\qquad\qquad
\left\{
\begin{array}{l}
\rho = 0\\
B_1 = 0\\
\partial_2 B_2 = 0\\
\eta = 0\\
\zeta = 0
\end{array}
\right.\,,
\label{eq:bcir}
\ee
and
\be
z=z_{\UV}\ :
\quad
\left\{
\begin{array}{l}
\phi = -i\\
A_1 = 0\\
\partial_2 A_2 = 0\\
s=0\\
\lambda = 2 i M_q \left(z_{\UV}/z_{\IR}\right)^{\Delta_-}
\end{array}
\right.
\qquad\qquad
\left\{
\begin{array}{l}
\chi = i\\
w = -1\\
Q = 0
\end{array}
\right.
\qquad\qquad
\left\{
\begin{array}{l}
\rho = 0\\
B_1 = 0\\
\partial_2 B_2 = 0\\
\eta = 0\\
\zeta = 0
\end{array}
\right.\,.
\label{eq:bcuv}
\ee

The boundary conditions at $r=\infty$ are obtained by the requirement
that the energy of the solution minus the vacuum energy for the scalar field is finite.
To obtain a soliton solution with $B=1$ one imposes
\be
r=\infty\ :
\quad
\left\{
\begin{array}{l}
\phi = -i e^{i \pi (z-z_{\UV})/(z_{\IR}-z_{\UV})}\\
\partial_1 A_1 = 0\\
A_2 = \displaystyle \pi/(z_{\IR}-z_{\UV})\\
s=0\\
\lambda = 2 i v(z) e^{i \pi (z-z_{\UV})/(z_{\IR}-z_{\UV})}
\end{array}
\right.
\quad\quad
\left\{
\begin{array}{l}
\chi = i e^{i \pi (z-z_{\UV})/(z_{\IR}-z_{\UV})}\\
w = -1\\
Q = 0
\end{array}
\right.
\quad\quad
\left\{
\begin{array}{l}
\rho = 0\\
\partial_1 B_1 = 0\\
B_2 = 0\\
\eta = 0\\
\zeta = 0
\end{array}
\right.\,.
\label{eq:bcrinfty}
\ee

On the $r=0$ boundary of the domain we must require the 2d solution
to give rise to regular 5d fields
and the gauge choice in eq.~(\ref{eq:2Dgauge}) to be fulfilled. These conditions are
fulfilled with the choice
\be
r=0\ :
\quad
\left\{
\begin{array}{l}
\phi_1/r \rightarrow A_1\\
(1+\phi_2)/r \rightarrow 0\\
A_2 = 0\\
\partial_1 A_1 = 0\\
s=0\\
\lambda_1 = 0\\
\partial_1 \lambda_2 = 0
\end{array}
\right.
\quad\qquad
\left\{
\begin{array}{l}
\chi_1 = 0\\
\partial_1 \chi_2 = 0\\
w = -\chi_2\\
Q = 0
\end{array}
\right.
\qquad\qquad
\left\{
\begin{array}{l}
\rho/r \rightarrow B_1\\
\partial_1 B_1 = 0\\
B_2 = 0\\
\eta_1 = \zeta\\
\eta_2 = 0\\
\partial_1 \zeta = 0
\end{array}
\right.\,.
\label{eq:bcr0}
\ee

\section{The QCD Anomaly} \label{app:Anomaly}

In this appendix we describe different forms of the QCD anomaly and discuss the relation with
the CS term included in the 5d theory. Part of the material that follows overlaps with
appendix A of \cite{Becciolini:2009fu}.

The CS term (\ref{Scs}) can be written as
\be
S_{CS} = -\frac{N_c}{24 \pi^2} \int d^5x \left[\ov\omega_5({\bf L}) - \ov\omega_5({\bf R})\right]\,,
\ee
where $\ov\omega_5$ differs by a total differential from the standard text-book CS form:
\begin{eqnarray}
 \omega_5({\mathbf A}) & = &  \ii\,\tr\left[\A\,\left(\dd \A\right)^2\,+\,\frac32 \A^3\dd\A\,+\,\frac35\A^5\right]\nn\\
&=& \ii\,\frac{1}{4}\, \hat{A} (\dd \widehat{A})^2 + \ii\frac{3}{2}\, \widehat{A} \tr\left[F^2\right]
+\ii\,\dd\left[\widehat{A}\,\tr\left[A\,F - \frac14 A^3\right]\right]\,\equiv\,{\overline{\omega}}_5({\mathbf A})\,+\,\dd X({\bf A})
\,.\hspace{2em}
\end{eqnarray}
The variation of the CS is given by eq.~(\ref{eq:DeltaS}), where the 4-form
\beq
\displaystyle
{\overline\omega}_{4}^1({\boldsymbol\alpha},{\bf A})\,=\,
\frac14{\widehat{\alpha}}\left(\dd{\widehat{A}}\right)^2\,+\,
\frac32{\widehat{\alpha}}\,\tr\left[F^2\right]\,,
\label{ombar}
\eeq
is defined from the relation
$\delta_{\alpha}{\overline\omega}_5=\dd{\overline\omega}_{4}^1$
and it is related to the standard $\omega_4^1$ by
\be
\ov\omega_4^1({\boldsymbol \alpha}, {\bf A}) = \omega_4^1({\boldsymbol \alpha}, {\bf A}) - \delta_\alpha X({\bf A})\,,
\ee
where
\beq
\displaystyle
{\omega}_{4}^1({\boldsymbol\alpha},{\mathbf A})\,=\,\tr\left[{\boldsymbol\alpha}\,\dd\left(\A\dd \A+\frac12 \A^3\right)\right]\,.
\eeq

Provided the IR term in eq.~(\ref{eq:DeltaS}) is cancelled, the CS variation gives the
anomaly of eq.~(\ref{eq:Anomaly}), which however does not coincide with the standard
text-book QCD anomaly that is normally put in the ``symmetric'' form
\beq
{\mathcal A}_{sym}\,=\,\frac{N_c}{24\pi^2}\int\left[{\omega}_{4}^1({\boldsymbol\alpha}_L,{\mathbf l})
-{\omega}_{4}^1({\boldsymbol\alpha}_R,{\bf r})
\right] \,.
\label{anst}
\eeq
The two forms of the anomaly (${\mathcal A}$ and ${\mathcal A}_{sym}$) are equivalent
because they only differ by a local counterterm
\be
{\mathcal A} = {\mathcal A}_{sym} - \frac{N_c}{24 \pi^2} \left[\delta_{{\boldsymbol \alpha}_L} X({\bf L})
- \delta_{{\boldsymbol \alpha}_R} X({\bf R})\right]\,.
\ee
Obviously we may equivalently have added the $X$ local counterterm to the 5d Lagrangian and kept
the standard form of the QCD anomaly. This would have not affected any of our results because $X$
only depends on the 4d ${\bf l}$ and ${\bf r}$ sources, whose physical value is zero.

Similarly, the QCD anomaly could be also put in the Adler--Bardeen form. Starting from
the symmetric anomaly, this is achieved by the addition of the Adler--Bardeen counterterm,
as explained in \cite{Panico:2007qd}.


\end{document}